\documentclass[12pt,reqno]{amsart}
\usepackage[utf8]{inputenc}
\usepackage[main=english,russian]{babel}%Russian for one of the references
\usepackage{amsfonts,amssymb,amsmath,mathtools,braket,dsfont,empheq,stmaryrd}
\usepackage{graphicx}
\usepackage{enumitem}

\theoremstyle{plain}
\newtheorem{theorem}{Theorem}%[section]
\newtheorem{lemma}[theorem]{Lemma}
\newtheorem{corollary}[theorem]{Corollary}
\newtheorem{proposition}[theorem]{Proposition}

\newtheorem{definition}[theorem]{Definition}
\newtheorem{assumption}[theorem]{Assumption}
\newcommand\xqed[1]{%
	\leavevmode\unskip\penalty9999 \hbox{}\nobreak\hfill\quad\hbox{#1}%
}
\newcommand\remarkend{\xqed{$\triangle$}}
\makeatletter
\def\@endtheorem{\remarkend\endtrivlist\@endpefalse }
\makeatother

\theoremstyle{remark}
\newtheorem{remark}[theorem]{Remark}
\newtheorem*{remark*}{Remark}
\theoremstyle{notation}

\newtheorem*{notation*}{Notation}
\makeatletter
\def\@endtheorem{\endtrivlist\@endpefalse }
\makeatother
\newcommand{\Par}[1]{\left( #1 \right)}
\newcommand{\Ee}[1]{\exp\Par{#1}}
\newcommand{\di}{\,\mathrm{d}}
\renewcommand{\Re}{\operatorname{Re}}

\renewcommand{\leq}{\leqslant}
\renewcommand{\geq}{\geqslant}

\newcommand{\N}{\mathbb{N}} % Natural numbers
\newcommand{\Cbb}{\mathbb{C}}%Can't use "\C" because Russian babel or T2A already uses it...
\newcommand{\R}{\mathbb{R}} % Real numbers

\newcommand{\cH}{\mathcal{H}}
\newcommand{\cA}{\mathcal{A}}

\newcommand{\pscalDISPLAY}[2]{{\left\langle #1, #2 \right\rangle}} %Scalar product automatically correctly sized
\newcommand{\pscalINLINE}[2]{{\langle #1, #2 \rangle{\vphantom{\pscalDISPLAY{1}{1}}}}} %Scalar product automatically correctly sized
\newcommand{\pscal}[2]{{\mathchoice{\pscalDISPLAY{#1}{#2}}{\pscalINLINE{#1}{#2}}{\pscalINLINE{#1}{#2}}{\pscalINLINE{#1}{#2}}}}%The four parameters of \mathchoice are, in order, for display, text, script, and scriptscript cases. And extra {} is needed around \mathchoice to avoid spacing issues.

\newcommand{\normDISPLAY}[1]{{\left\vert\kern-0.25ex\left\vert #1 \right\vert\kern-0.25ex\right\vert}}
\newcommand{\normINLINE}[1]{{\vert\kern-0.25ex\vert #1 \vert\kern-0.25ex\vert{\vphantom{\normDISPLAY{1}}}}}
\newcommand{\norm}[1]{{\mathchoice{\normDISPLAY{#1}}{\normINLINE{#1}}{\normINLINE{#1}}{\normINLINE{#1}}}}%The four parameters of \mathchoice are, in order, for display, text, script, and scriptscript cases. And extra {} is needed around \mathchoice to avoid spacing issues.
\newcommand{\normSM}[1]{\normINLINE{#1}}

\newcommand*{\transp}{^{\mkern-1.5mu\mathsf{T}}}
\renewcommand{\le}{\leq}
\renewcommand{\ge}{\geq}
\newcommand{\bigO}{\mathcal{O}}
\DeclareMathOperator{\Id}{\mathds{1}}
\DeclareMathOperator{\sgn}{sign}
\DeclareMathOperator{\vect}{span}
\DeclareMathOperator{\ac}{AC}
\DeclareMathOperator{\odd}{odd}
\DeclareMathOperator{\even}{even}

\usepackage{xspace} % for \NLS command
\newcommand{\NLS}{\relax\ifmmode \mathrm{(NLS)}\else\textup{(NLS)}\xspace\fi}
\newcommand{\NLD}{\relax\ifmmode \mathrm{(NLD)}\else\textup{(NLD)}\xspace\fi}
\newcommand{\oGP}{\relax\ifmmode \mathrm{(oGP)}\else\textup{(oGP)}\xspace\fi}
\newcommand{\noEV}{\relax\ifmmode \mathrm{(noEV)}\else\textup{(noEV)}\xspace\fi}
\newcommand{\noRes}{\relax\ifmmode \mathrm{(noRes)}\else\textup{(noRes)}\xspace\fi}

\usepackage[table, dvipsnames]{xcolor}
\usepackage{tikz}
\newcommand{\absenceSpectrum}{Red}
\newcommand{\absenceSpectrumnew}{Cyan}
\newcommand{\existingSpectrum}{Black}
\newcommand{\possibleSpectrum}{Black}
\usepackage{float}
\usepackage{graphicx}
\usepackage{subcaption}

\usepackage{nicefrac}		
\usepackage{hyperref}

\title[On gap properties for the linearized 1D Dirac--Soler model]{On gap properties\\for the linearized 1D Dirac--Soler model}

\author[D. Aldunate]{Danko Aldunate}
\address{Centro de Modelamiento Matemático (CNRS IRL2807), Universidad de Chile, Santiago, Chile}
\email{dmaldunate@cmm.uchile.cl}

\author[J. Ricaud]{Julien Ricaud}
\address{Departamento de F\'{i}sica Matem\'{a}tica, Instituto de Investigaciones en Matem\'aticas Aplicadas y en Sistemas, Universidad Nacional Aut\'onoma de M\'exico, CP 04510, Ciudad de M\'exico, M\'exico}
\email{ricaud@iimas.unam.mx}

\author[E. Stockmeyer]{Edgardo Stockmeyer}
\address{Instituto de F\'isica, Pontificia Universidad Cat\'olica de Chile, Vicu\~na Mackenna 4860, Santiago 7820436, Chile.}
\email{stock@uc.cl}

\date{\today}

\allowdisplaybreaks

\begin{document}
	
\begin{abstract}
We study spectral properties of the Dirac operator $L_0$ obtained by linearizing the one-dimensional Soler model around standing waves with power nonlinearity $f(s)=s|s|^{p-1}, \, p>0$.
We give a sharp characterization of the spectral gap. More precisely, we prove that if $p\ge1$, then the gap contains no eigenvalues other than the symmetry-induced energies $-2\omega$ and $0$, while for~$0<p<1$ additional eigenvalues bifurcate from the thresholds of the essential spectrum and enter the gap. We further show that the thresholds are never eigenvalues for any $p>0$, and that threshold resonances are absent for~$p>1$. Along the way, we establish several structural results for generalized eigenfunctions of one-dimensional Dirac operators at threshold energies under mild assumptions on the potential.
\end{abstract}
	
\maketitle
\tableofcontents
\section{Introduction}

We study the spectral structure of one-dimensional Dirac operators arising from the linearization of solitary waves for the nonlinear Dirac equation with Soler nonlinearity. More precisely, we investigate the presence or absence of eigenvalues inside the spectral gap, as well as the possible occurrence of resonances at the thresholds of the essential spectrum. These spectral questions are closely connected to the stability and long-time dynamics of standing waves.

Let $m>0$ and consider the one-dimensional free Dirac operator
\[
	D_m := i\sigma_2\partial_x + m\sigma_3
\]
acting on~$L^2(\R,\Cbb^2)$, where
\[
	\sigma_1=
	\begin{pmatrix}
		0&1\\
		1&0
	\end{pmatrix},
	\qquad
	\sigma_2=
	\begin{pmatrix}
		0&-i\\
		i&0
	\end{pmatrix},
	\qquad
	\sigma_3=
	\begin{pmatrix}
		1&0\\
		0&-1
	\end{pmatrix}.
\]

We consider the nonlinear Dirac equation in~$1+1$ dimensions with Soler-type nonlinearity~\cite{Ivanenko-38-PZS,Soler-70} and initial data $\phi \in H^1(\R, \Cbb^2)$,
\begin{equation}\label{NLDE}
	\left\{
	\begin{aligned}
		i\partial_t \psi &= D_m \psi - f\!\left(\pscal{\psi}{\sigma_3 \psi}_{\Cbb^2}\right)\sigma_3\psi\,, \\
		\psi(\cdot,0) &= \phi\,,
	\end{aligned}
	\right.
\end{equation}
where $\psi=(\psi_1,\psi_2)\transp:\R\times\R\to\Cbb^2$. For a broad class of nonlinearities~$f$, this equation admits standing waves of the form (see, e.g.,~\cite{CazVaz-86,BerCom-12,AldRicStoVDB-23})
\[
	\psi(x,t) = e^{-i\omega t}\phi_0(x)\,, \qquad \omega\in(0,m)\,.
\]
Here, $\phi_0\in H^1(\R,\mathbb{C}^2)$ is a localized profile. By symmetry, the case $\omega\in(-m,0)$ can be treated analogously. We also refer to~\cite{AlvSol-83,BouCuc-12,PelSte-12,ComGuaGus-14,PelShi-14,BerComSuk-15,BouCom-16,BouCom-17,Lakoba-18,BouCom-19-Book,BouCom-19,BouCacCarComNojPos-20-arXiv,BouComKul-24-arXiv} for further results on existence and stability of standing waves.

The profile $\phi_0$ satisfies the stationary equation
\begin{equation}\label{Stationary_equation_L0}
	\left( D_m - f\!\left(\pscal{\phi_0}{\sigma_3\phi_0}_{\Cbb^2}\right)\sigma_3 - \omega \right)\phi_0 = 0\,.
\end{equation}
We denote the associated self-adjoint operator by
\[
	L_0 := D_m - f\!\left(\pscal{\phi_0}{\sigma_3\phi_0}_{\Cbb^2}\right)\sigma_3 - \omega\Id\,.
\]

More generally, we consider the family of self-adjoint operators
\begin{equation}\label{Def_L_mu}
	L_\mu \equiv L_\mu(\omega) := L_0 - \mu Q\,, \qquad \mu\in\R\,,
\end{equation}
with domain~$H^1(\R,\Cbb^2)$, where
\begin{equation}\label{Def_of_Q}
	Q := f'\!\left(\pscal{\phi_0}{\sigma_3\phi_0}_{\Cbb^2}\right) (\sigma_3\phi_0)(\sigma_3\phi_0)\transp\,.
\end{equation}

Unless stated otherwise, we consider the power nonlinearity
\begin{align}\label{nl}
	f(s)=s|s|^{p-1}\,, \qquad p>0\,,
\end{align}
and denote by $\phi_0\equiv\phi_{0,p}\equiv(v_p,u_p)\transp$ the standing-wave profile corresponding to the exponent $p$. If there is no risk of confusion we simply write $L_\mu\equiv L_\mu(p,\omega)$ .

The operators $L_0$ and $L_2$ arise as the off-diagonal components of the linearized operator
\begin{equation}\label{Def_H}
	H :=
	\begin{pmatrix}
		0 & L_0 \\
		L_2 & 0
	\end{pmatrix}
\end{equation}
obtained by linearization around the standing wave $e^{-i\omega t}\phi_0$; see~\cite{AldRicStoVDB-23} for the conventions adopted here. With this convention, spectral stability of the standing wave corresponds to the reality of the spectrum of~$H$. 
Consequently, understanding the spectra of~$L_0$ and~$L_2$ is a natural and important step toward the stability theory of the one-dimensional Soler model.

For the operators $L_\mu$ defined above, the essential spectrum is given by
\[
\sigma_{\mathrm{ess}}(L_\mu)
=
(-\infty,-m-\omega]\cup[m-\omega,+\infty)\,.
\]

The spectral questions addressed in this paper are motivated by the corresponding theory for nonlinear Schr\"odinger equations. In the \NLS setting, a fundamental issue is whether the linearized operator possesses additional discrete spectrum between the neutral modes generated by symmetries and the threshold of the essential spectrum, and whether resonances occur at that threshold. These properties are often grouped under the name \emph{gap property}, either as a structural assumption or as a theorem to be proved, depending on the model under consideration. They play a central role in orbital and asymptotic stability theory, since they govern the interaction between discrete modes and the continuous spectrum, and more generally influence the long-time dynamics; see, for example,~\cite{DemSch-06,Schlag-09,NakSch-11}.

The gap property has been established for one-dimensional \NLS in~\cite[Lemma~9.1 and Proposition~9.2]{KriSch-06}, and for the cubic case in three dimensions in~\cite[Theorem~1.1]{LiYan-26}; see also~\cite{ChaGusNakTsa-08,CosHuaSch-11} and the references therein.

%%%
Motivated by this framework, we investigate the analogous spectral questions for the linearized Dirac operators considered here. Unlike the Schrödinger case, the Dirac operator is not bounded from below. Moreover, it possesses two threshold energies together with a symmetry-induced pair of internal eigenvalues, leading to a substantially richer spectral structure.

Owing to the $\sigma_1$-symmetry of the spectrum of~$L_0+\omega$ around zero, the operator $L_0$ has two distinguished eigenvalues, namely $-2\omega$ and $0$; see, for instance,~\cite[Proposition~1.2 and Theorem~1.3]{AldRicStoVDB-23}. These correspond to the explicit eigenstates $\sigma_1\phi_0$ and $\phi_0$, respectively, and play the role of the distinguished internal modes around which the spectral gap is organized. These are analogous to the distinguished ground states in the Schr\"odinger setting. Moreover, it was proved in~\cite{AldRicStoVDB-23} that
\[
	\sigma(L_0)\cap(-2\omega,0)=\emptyset
\]
for all $p>0$, and that all eigenvalues of~$L_\mu$ are simple.

As a consequence, the Dirac analogue of the gap property naturally decomposes into three distinct spectral properties: For $\mu\in\{0,2\}$,
\begin{itemize}[leftmargin=5em]
	\item[\oGP]\label{oGP} \emph{open gap property:}
		\[
	\sigma(L_\mu)\cap\bigl((-m-\omega,-2\omega)\cup(0,m-\omega)\bigr)=\emptyset\,;
	\]
	
	\item[\noEV]\label{noEV} $L_\mu$ has no eigenvalues at the thresholds of its essential spectrum;
	
	\item[\noRes]\label{noRes} $L_\mu$ has no resonances at the thresholds of its essential spectrum.
\end{itemize}
\bigskip
The main results of this paper reveal a sharp spectral transition at the
critical exponent $p=1$. In the Gross--Neveu case $p=1$~\cite{GroNev-74},
the operator $L_0$ possesses threshold resonances. For $0<p<1$, these
resonances bifurcate into genuine eigenvalues inside the spectral gap,
while for~$p>1$ no threshold resonances occur. In this sense, resonances
at the thresholds generate eigenvalues in the spectral gap under variation
of the nonlinearity exponent. This spectral transition is illustrated in
Figure~\ref{fig:spectral_transition}.

\begin{figure}[ht]
    \centering
	\pgfmathsetmacro{\M}{2}          % masa m
\pgfmathsetmacro{\W}{0.6}          % omega, en (0,\M)  <-- lo unico que hay que tocar
\pgfmathsetmacro{\Tp}{\M-\W}     % umbral superior   m-omega
\pgfmathsetmacro{\Tm}{-\M-\W}    % umbral inferior  -m-omega
\pgfmathsetmacro{\Gs}{-2*\W}     % -2 omega
\pgfmathsetmacro{\Cu}{0.75*\Tp}  % curvatura de las ramas
\pgfmathsetmacro{\bU}{\Tp+0.3}   % borde exterior de la banda superior
\pgfmathsetmacro{\bL}{\Tm-0.4}   % borde exterior de la banda inferior

\begin{tikzpicture}[x=5.5cm, y=2.5cm,
	    eje/.style = {-latex },
	    known/.style = {very thick, black},
	    new/.style = { very thick, \possibleSpectrum!60, dashed},
	    newCircle/.style = { very thick, black, dotted}, yscale=0.7,
	    res/.style    = {circle, draw=black, fill=white, line width=0.9pt, inner sep=0pt, minimum size=5pt},
	    reshyp/.style = {circle, draw=black, fill=white, line width=0.5pt, inner sep=0pt, minimum size=4pt}]

	%% ESS-Spect
	    \shade[left color=\existingSpectrum!60,right color=\existingSpectrum!10, shading angle =180]
	        (0,\Tp) -- (2,\Tp) -- (2,\bU) -- (0,\bU) --cycle;
	    \shade[left color=\existingSpectrum!10,right color=\existingSpectrum!60, shading angle =180]
	        (2,\bL) -- (0,\bL) -- (0,\Tm) -- (2,\Tm) --cycle;

	%%AXES
	    \draw[eje](0,{\bL+0.1})--(0,{\bU+0.1}) node [right] {\footnotesize $\sigma(L_0 (p))$};
	    \draw[eje](0,\bL) --(2,\bL) node[below]{\footnotesize $p$} ;

	%% EIGENVALUES
	    \draw[known](0,0) --(2,0) ;
	    \draw[known](0,\Gs) --(2,\Gs) ;
	    \shade[left color=black, right color=.!30, shading angle=0]
    		plot[domain=0:1, samples=100] (\x, {\Tp - \Cu*(1-\x)^2 + 0.01})
    			-- (1,\Tp)
    			-- plot[domain=1:0, samples=100] (\x, {\Tp - \Cu*(1-\x)^2 - 0.01})
    			-- cycle;
	    \shade[left color=black, right color=.!20, shading angle=0]
    		plot[domain=0:0.7, samples=100] (\x, {\Tp - \Cu*(1 - (\x + 0.3))^2 + 0.005})
    			-- (0.7,\Tp)
    			-- plot[domain=0.7:0, samples=100] (\x, {\Tp - \Cu*(1 - (\x + 0.3))^2 - 0.005})
    			-- cycle;
	    \shade[left color=black, right color=.!25, shading angle=0]
    		plot[domain=0:0.3, samples=100] (\x, {\Tp - \Cu*(1 - (\x + 0.7))^2 + 0.005})
    			-- (0.3,\Tp)
    			-- plot[domain=0.3:0, samples=100] (\x, {\Tp - \Cu*(1 - (\x + 0.7))^2 - 0.005})
    			-- cycle;

	%% NEGATIVE EIGENVALUES  (reflejo: y -> -2w - y)
	    \shade[left color=.!30, right color=black, shading angle=0]
    		plot[domain=0:1, samples=100] (\x, {\Tm + \Cu*(1-\x)^2 + 0.01})
    			-- (1,\Tm)
    			-- plot[domain=1:0, samples=100] (\x, {\Tm + \Cu*(1-\x)^2 - 0.01})
    			-- cycle;
	    \shade[left color=.!20, right color=black, shading angle=0]
    		plot[domain=0:0.7, samples=100] (\x, {\Tm + \Cu*(1 - (\x + 0.3))^2 + 0.005})
    			-- (0.7,\Tm)
    			-- plot[domain=0.7:0, samples=100] (\x, {\Tm + \Cu*(1 - (\x + 0.3))^2 - 0.005})
    			-- cycle;
	    \shade[left color=.!25, right color=black, shading angle=0]
    		plot[domain=0:0.3, samples=100] (\x, {\Tm + \Cu*(1 - (\x + 0.7))^2 + 0.005})
    			-- (0.3,\Tm)
    			-- plot[domain=0.3:0, samples=100] (\x, {\Tm + \Cu*(1 - (\x + 0.7))^2 - 0.005})
    			-- cycle;

	    \draw[dotted,thick] (1,\bL) -- (1,\Tp);

	%% RESONANCES
	    \node[res,fill=orange]    at (1,\Tp)   {};   % p=1
	    \node[reshyp] at (0.7,\Tp) {};   % hipotetica
	    \node[reshyp] at (0.3,\Tp) {};   % hipotetica
	    \node[res,fill=orange]    at (1,\Tm)   {};
	    \node[reshyp] at (0.7,\Tm) {};
	    \node[reshyp] at (0.3,\Tm) {};

	%%LABELS
	    \draw[shift={(-4pt,0)}]    (0,0) -- (0pt,-0pt) node[] {\tiny$0$};
	    \draw[shift={(-10pt,\Gs)}] (0,0) -- (0pt,-0pt) node[] {\tiny$-2\omega$};
	    \draw[shift={(0,\Tp)}]     (0,0) -- (0,0) node[left] {\tiny$m-\omega$};
	    \draw[shift={(0,\Tm)}]     (0,0) -- (0,0) node[left] {\tiny$-m-\omega$};
	    \foreach \p in {0,1}{\draw[shift={(\p,\bL)}] (0,2pt) -- (0,-2pt) node[yshift=-4pt] {\tiny$\p$};}
\end{tikzpicture}	

	\captionsetup{width=.9\textwidth}
	\caption{Sketch of the spectrum of $L_0$ as a function of $p$ and $\omega$ fixed. Lines: branches of eigenvalues; Disks: resonances.\protect\\
	Orange disk: known threshold resonance at $p=1$; White disk: possible resonances for $p<1$ (whose existence can be found numerically).\protect\\
		Thick lines: branches of eigenvalues whose existence has been established. Namely, $0$, $-2\omega$ and the branch, for $p \in (0,1)$, to which the resonance at~$p=1$ gives birth, as established in Theorem~\ref{main_thm_1_gap_p_below_1} and Proposition~\ref{prop-birth}.\protect\\
		Thinner lines: sketch of other branches of eigenvalues whose birth occurs at any $p<1$ such that there exists a resonance for this $p$, as established in Proposition~\ref{prop-birth}.\protect\\
		Except for the known eigenvalues $0$ and $-2\omega$, the other branches are sketched fading as~$p \to 0$ in order to indicate that we do not know their shape and even less to what they converge as~$p \to 0$. In particular, it is \emph{not} a way to sketch that they would disappear from the spectrum for very small $p$.\protect\\
		}
		\label{fig:spectral_transition}
\end{figure}

More precisely, we prove that the open gap property~\oGP holds for~$L_0$
if and only if $p\ge1$, and that the threshold energies are never
eigenvalues of~$L_0$ for any $p>0$. We further show that threshold
resonances are absent for all $p>1$, whereas for~$p=1$ they are known to
exist; see, for instance,~\cite[Lemma~5.5]{BerCom-12} and
\cite[Lemma~6.1]{AldRicStoVDB-23}. Finally, we establish the general
simplicity result that each threshold supports at most one generalized
eigenfunction. 

For $\mu\neq0$, the open gap property may fail even in regimes where it
holds for~$L_0$; see~\cite{AldRicStoVDB-23}, Corollary~\ref{main_cor} and the comments at the end of this section.

\subsection{Main results.}
Throughout the statements below, let $L_\mu\equiv L_\mu(p,\omega)$ be
as in~\eqref{Def_L_mu} with nonlinearity $f(s)=s|s|^{p-1}$.

We begin by characterizing the discrete spectrum inside the spectral gap of~$L_0$.
\begin{theorem}[Characterization of eigenvalues in the gap]\label{main_thm_1_gap}
	Let $(p,\omega)\in(0,+\infty)\times(0,m)$.
	\begin{enumerate}[label=(\arabic*), ref=\thetheorem(\arabic*)]
		\item\label{main_thm_1_gap_p_above_1} If $p\geq1$, then
		\[
			\sigma \left(L_0\right) \cap (-m-\omega, m-\omega) = \{-2\omega, 0\}\,.
		\]
		
		\item\label{main_thm_1_gap_p_below_1} If $p<1$, then $L_0$ admits at least two additional eigenvalues: $\lambda_+(p) \in (0,m-\omega)$ and $\lambda_-(p) \in (-m-\omega,0)$.
%		Moreover, by symmetry of the spectrum of~$L_0$ with respect to $-\omega$, there exists also an eigenvalue $\lambda_-(p) \in (-m-\omega,0)$ of~$L_0$ satisfying
%		\[
%			\lambda_-(p) = -m-\omega+\mathcal{O}\!\left((1-p)^{2+s}\right)\,, \qquad \forall\, s>0\,.
%		\]
	\end{enumerate}
\end{theorem}

\begin{remark*}\leavevmode
	\begin{enumerate}[label=\roman*)]
		\item The above result for the case $p=1$ was already given in \cite[Lemma~6.2]{AldRicStoVDB-23}.
		\item In Proposition~\ref{prop-birth} below, we establish asymptotic bounds on $\lambda_\pm(p)$ when $p \nearrow 1$.
		\item Note that Theorem~\ref{main_thm_1_gap_p_above_1} allows a direct improvement of the result in~\cite{AldRicStoVDB-23} about the range of~$\omega$'s for which $H$ defined in~\eqref{Def_H} has no non-zero eigenvalues on the imaginary axis. Indeed, combining it with~\cite[Fig.~5 \& Theorem~5.3]{AldRicStoVDB-23}, the previously known range of~$\omega$'s is increased, see Figure~\ref{Fig_Range_lower_bound_on_Re_z2}.
		\item By symmetry of the spectrum of~$L_0$ with respect to $-\omega$, the existence of~$\lambda_\pm$ immediately yields that of~$\lambda_\mp$.
	\end{enumerate}
\end{remark*}

Next, we rule out eigenvalues at the thresholds of~$L_0$.
\begin{theorem}[Absence of eigenvalues at the thresholds]\label{main_thm_2_noEV}
	Let $(p,\omega)\in(0,+\infty)\times(0,m)$. Then, the threshold energies $\pm m - \omega$ do not belong to the point spectrum of~$L_0$.
\end{theorem}
In order to address our next results let us introduce the notion of resonances in the more general framework of operators of the form $D_m+V$ under mild assumptions on~$V$.
\begin{definition}\label{Def_generalized_eigenfunction}
	Let $H=D_m+V$ with $V\in L^1_{loc} (\R,\Cbb^{2\times2})$ and domain~$D(H)\subset H^1 (\R)$.
	
	We say that $\Psi: \R \to\Cbb^2$ is a \emph{generalized eigenfunction} of~$H$ associated with a \emph{generalized eigenvalue} $\lambda$ if it is a non-zero $L^\infty(\R)$-distributional solution to~$H\Psi=\lambda\Psi$.
	
	If $\Psi$ is not square integrable we call it a \emph{resonance}.
\end{definition}
\begin{remark*}
	Note that for operators such that their $L^2$-eigenfunctions belong to~$H^1(\R)$, like~$L_\mu$, the space of generalized eigenfunctions contains exactly both eigenfunctions ($L^2 (\R)$-solutions) and resonances ($L^\infty (\R)\setminus L^2 (\R)$-solutions), since $H^1(\R) \subset L^\infty(\R)$.
\end{remark*}
The next theorem shows that the threshold resonances present in the Gross--Neveu case (Remark~\ref{rmres} below for their explicit form) disappear completely as soon as $p>1$.
\begin{theorem}[Absence of threshold resonances]\label{main_thm_noRes}
	Let $p>1$ and $\omega\in (0,m)$. Then, the operator $L_0(p,\omega)$ does not have resonances at its threshold energies $\pm m-\omega$.
\end{theorem}
For the operator $L_0$, the previous results already provide a rather detailed description of threshold phenomena. Indeed, Theorem~\ref{main_thm_2_noEV} excludes threshold eigenvalues for all $p>0$, while the theorem above shows that threshold resonances are absent for~$p>1$. Thus, for the operator $L_0$, the following result is mainly relevant in the regime $p\in(0,1]$.
	
Its main interest, however, lies in the broader framework of the operators $L_\mu$, $\mu\ge0$, and more generally for one-dimensional Dirac operators with rough potentials (see  Section~\ref{strategy} for more details).
\begin{theorem}[Simplicity of threshold generalized eigenvalues for $L_\mu$]\label{main_thm_3_Resonances}\leavevmode

	Let $(p,\omega)\in(0,+\infty)\times(0,m)$, $\mu\in\R$, $L_\mu$ be as in~\eqref{Def_L_mu}, and $\tau\in\{\pm1\}$. Then, 
	\[
		\dim\{\Psi\in L^\infty(\R,\mathbb{C}^2 ) : L_\mu\Psi=(\tau m-\omega)\Psi\}\le1\, ,
	\]
	where $\Psi$ solves the equation in the distributional sense. 
\end{theorem}
The next section outlines the main mechanisms underlying the proofs of our results. The proof of Theorem~\ref{main_thm_1_gap} relies on a rescaling identity for a family of auxiliary operators together with suitable applications of the min--max principle for operators with a spectral gap. This variational mechanism also yields the absence of threshold resonances for~$p>1$: any hypothetical resonance would generate eigenvalues inside the gap for nearby exponents $p>1$, contradicting Theorem~\ref{main_thm_1_gap_p_above_1}.

\bigskip

We close the present introduction by giving an immediate corollary of Theorem~\ref{main_thm_1_gap} for~$L_2$ defined in~\eqref{Def_L_mu}. This result extends the one established only in the non-relativistic $\omega \to m$ limit in~\cite[Corollary~7.4 \& Theorem~7.3]{AldRicStoVDB-23} and is a step in the study of the spectrum of the operator $L_2$ appearing in the linearized operator~\eqref{Def_H}.
\begin{corollary}\label{main_cor}
	Let $(p,\omega)\in(0,+\infty)\times(0,m)$.
	\begin{enumerate}
		\item If $p\geq 1$, then $\# \Par{ \sigma \left(L_2 \right) \cap (-m-\omega, -2\omega)} = 0$. 
		\item If $p<1$, 	then $\# \Par{ \sigma \left(L_2 \right) \cap (-2\omega,m-\omega)}\geq 3$ 
	\end{enumerate}
\end{corollary}
This corollary is obtained from Theorem~\ref{main_thm_1_gap}, Hellmann--Feynman theorem, and the simplicity of the eigenvalues of~$L_\mu$ proven in~\cite[Lemma~2.3]{AldRicStoVDB-23}. In particular, it shows that \oGP is not satisfied for~$L_\mu$ for all $0 < p < 1$ and $\mu \geq 0$. A similar result was obtained in \cite{AldRicStoVDB-23}, where the authors proved that $L_2$ exhibits new eigenvalues in~$(0,m-\omega)$~for all~$p > 0$~in the non-relativistic limit $\omega \to m$; see, for example, \cite[Corollary~7.4]{AldRicStoVDB-23}. Moreover, note that for any $p>0$ and $\omega \in (0,m)$, the operator $L_2$ has exactly one eigenvalue in~$(-2\omega,0)$, which does not contradict the \emph{gap condition}; see \cite[Theorem~7.1]{AldRicStoVDB-23}.

For the reader's convenience, in Figure~\ref{fig:spectra} we update Figure~1
from~\cite{AldRicStoVDB-23}, where the spectra of~$L_0$ and~$L_2$ are
sketched for general nonlinearities. Here, we restrict the sketch to power
nonlinearities with $p\geq 1$. In particular, our contribution to the
updated figure consists of proving the absence of eigenvalues other than
$0$ and $-2\omega$ for~$L_0$ when $p>1$, as well as the absence of
eigenvalues in~$(-m-\omega,-2\omega)$ for~$L_2$. As mentioned before, the
absence of eigenvalues for~$L_0$ other than the distinguished eigenvalues
$0$ and $-2\omega$ when $p=1$ was proved in~\cite{AldRicStoVDB-23}. Finally, in the case $p<1$, which we do not sketch, we identify new
eigenvalues in~$(-m-\omega,-2\omega)$ and $(0,m-\omega)$ for~$L_0$, and
in~$(0,m-\omega)$ for~$L_2$.
	\begin{figure}[ht]
	\centering
	%Sketch L_0
	\begin{tikzpicture}[scale=1.5,
	  eje/.style = {-latex },
	  known/.style = {ultra thick,\existingSpectrum},
	new/.style = { thick, violet}]
	%% ESS-Spect
	  \shade[right color=\existingSpectrum!60,left color=\existingSpectrum!10] (-1,0) -- (-2,1) --(-2.2, 1) -- (-2.2,0)--cycle;
	  \shade[left color=\existingSpectrum!60,right color=\existingSpectrum!10] (1,0) -- (0,1) --(1.9, 1) -- (1.9,0)--cycle;
	%% No-spectrum
	   \fill[color=\absenceSpectrumnew!25] (0,0) -- (-1,0) --(-2,1)--(0,1)--(1,0)--cycle;  
	  \fill[color=\absenceSpectrum!15] (0,0)--(-2,1)--(0,1)--cycle; 
	%%AXES
	  \draw[eje](-2.25, 0)--(2,0) node [below] {\footnotesize $\sigma(L_0(\omega))$};
	  \draw[eje](0,-0.2) --(0,1.2) node[right]{\footnotesize $\omega$} ;
	%% EIGENVALUES
	  \draw[known](0,0) --(0,1);
	  \draw[known](0,0) --(-2,1);
	%%LABELS
	  \draw[shift={(1,0)}] (0pt,0pt) -- (0pt,-2pt) node[below] {\tiny$+m$};
	  \draw[shift={(-1,0)}] (0pt,0pt) -- (0pt,-2pt) node[below] {\tiny$-m$};
	  \draw[shift={(-4pt,-4pt)}] (0pt,0pt) -- (0pt,-0pt) node[] {\tiny$0$};
	  \draw[shift={(0,1)}] (2pt,0) -- (-2pt,0) node[xshift=-4pt, yshift=4pt] {\tiny$m$}; 
	\end{tikzpicture} 
	%
	%Sketch L_2
	\begin{tikzpicture}[scale=1.5,
	  eje/.style = {-latex },
	  known/.style = {ultra thick,\existingSpectrum},
	new/.style = { thick, violet}]
	
	%% ESS-Spect
	  \shade[right color=\existingSpectrum!60,left color=\existingSpectrum!10] (-1,0) -- (-2,1) --(-2.2, 1) -- (-2.2,0)--cycle;
	  \shade[left color=\existingSpectrum!60,right color=\existingSpectrum!10] (1,0) -- (0,1) --(1.9, 1) -- (1.9,0)--cycle;
	%% No-spectrum
	   \fill[color=\absenceSpectrum!15] (0,0) -- (-2,1) --(0,1)--cycle;  
	  \fill[color=\absenceSpectrumnew!25] (0,0) -- (-1,0) --(-2,1)--cycle;  
	%%AXES 
	  \draw[eje](-2.25, 0)--(2,0) node [below] {\footnotesize $\sigma(L_2(\omega))$};
	  \draw[eje](0,-0.2) --(0,1.2) node[right]{\footnotesize $\omega$} ;
	%% EIGENVALUES
	  \draw[known](0,0) --(0,1) ;
	  \draw[known](0,0) --(-2,1) ;
	  \draw[known] (0,1)..controls(-3/8,0.66)..(0,0);
	%%LABELS
	  \draw[shift={(1,0)}] (0pt,0pt) -- (0pt,-2pt) node[below] {\tiny$+m$};
	  \draw[shift={(-1,0)}] (0pt,0pt) -- (0pt,-2pt) node[below] {\tiny$-m$};
	  \draw[shift={(-4pt,-4pt)}] (0pt,0pt) -- (0pt,-0pt) node[] {\tiny$0$};
	  \draw[shift={(0,1)}] (2pt,0) -- (-2pt,0) node[xshift=-4pt, yshift=4pt] {\tiny$m$}; 
\end{tikzpicture}

	\captionsetup{width=.9\textwidth}
	\caption{Spectra of the families of self-adjoint operators~$\{L_0(\omega)\}_\omega$ and~$\{L_2(\omega)\}_\omega$ with $f(s)=s|s|^{p-1}$ and $p\geq1$. The spectrum of one operator is the intersection of the sketch with a horizontal line. Shading gray: essential spectrum; solid black: known eigenvalues; salmon: region where the absence of eigenvalues was proved in \cite{AldRicStoVDB-23}; light blue: region where we prove the absence of eigenvalues.}
		\label{fig:spectra}
\end{figure}

\begin{figure}[ht]
	\includegraphics[width=0.8\columnwidth]{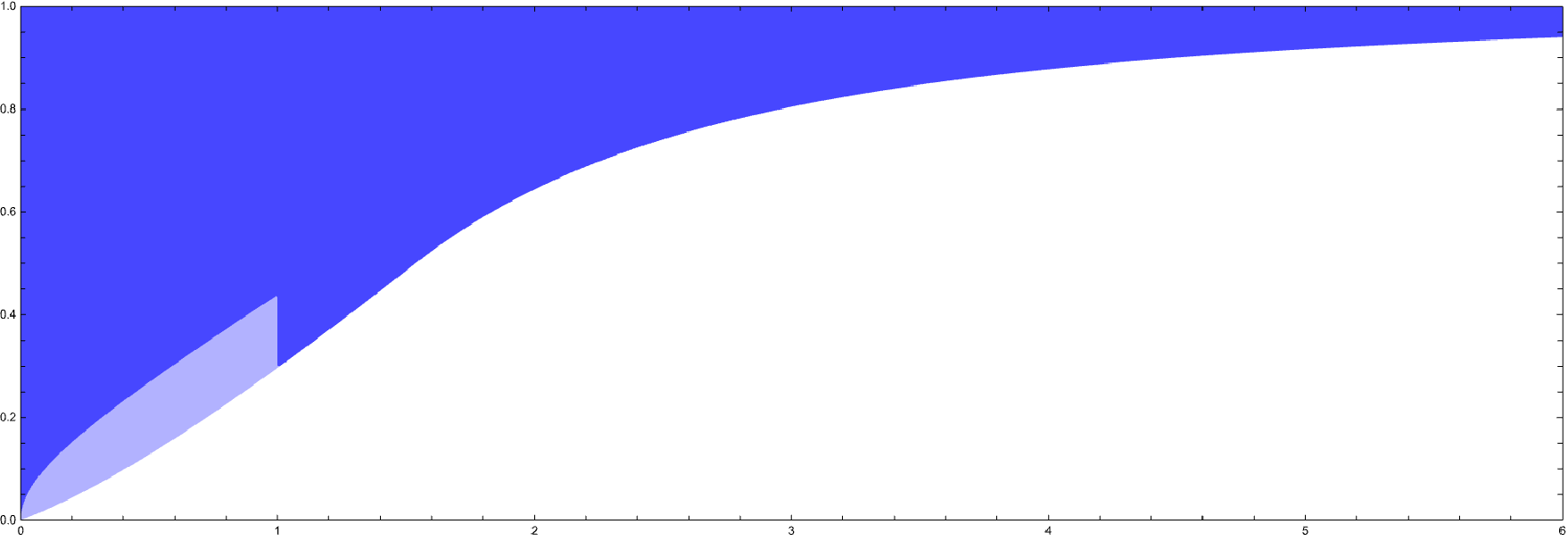}
	\captionsetup{width=.9\textwidth}
	\caption{Dark blue: range for which we have the \emph{cone condition} $\Re z^2 \geq 0$ for eigenvalues $z$ of $H_2$ outside the imaginary axis (see~\cite{AldRicStoVDB-23}). 
	Light blue: area where the real frontier above which the \emph{cone condition} $\Re z^2 \geq 0$ is satisfied lies; this exact frontier will depend on the smallest positive eigenvalue of $L_0$ thence it will be close to the bottom of this region for $p$ (below and) close to $1$ and close to its top for $p$ close to $0$.}
	\label{Fig_Range_lower_bound_on_Re_z2}
\end{figure}

\begin{notation*}
	We use $\norm{\cdot}$ to denote the operator norm if the entry is a bounded operator and $\norm{\cdot}_q$ for the $L^q$-norm, $q\in[1,\infty]$, of a spinor (or a function). Moreover, if there is no risk of confusion, we may use $\norm{\cdot}$ to denote the $L^2$-norm if the entry is an $L^2$-vector.
\end{notation*}

\bigskip

\textbf{Acknowledgments:} The authors are grateful to N. Boussaïd, A. Comech, J.-C. Cuenin, and H. Van~Den~Bosch for the interesting and fruitful discussions. D.A. and J.R.~thank N. Boussaïd and H. Van~Den~Bosch for pointing out, respectively, works on the \emph{gap property} for NLS and Kneser's criteria.
D.A.~acknowledges support from Centro de Modelamiento Matemático (CMM) BASAL fund FB210005 for center of excellence from ANID-Chile.
J.R. is Fellow of \emph{Sistema Nacional de Investigadoras e Investigadores} and his research was partially supported by the DGAPA-UNAM's project PAPIIT IA102326.
E.S.~acknowledges support from Fondecyt (ANID, Chile) through the grants \#123–1539 and \#125–0596.

\section{Strategy of the proofs \& some structural results}\label{strategy}
This section describes the main mechanisms underlying the proofs of our
main
 results and collects several structural properties used throughout
the paper. The central idea is a 
variational analysis of the transition at
$p=1$, based on a rescaled family of operators $\{A_p\}_{p>0}$ and the
min--max principle in a spectral gap. We then describe the structure of
generalized eigenfunctions at the thresholds, which plays a key role both
in the variational construction and in the analysis of threshold
eigenvalues and resonances.

\subsection{Variational mechanism governing the spectral transition}
Let us start by describing a crucial ingredient in our analysis, inspired by~\cite[Section~5]{AldRicStoVDB-23}. 
We introduce the unitary rescaling operator
\[
	U_p\psi(x) := p^{-\frac{1}{2}}\psi(p^{-1} x)\,,
\]
and define the operator $A_p$, unitarily equivalent to $L_0(p)+\omega$, by
\begin{equation}\label{Def_Ap}
	A_p \equiv A_p(m,\omega) := U_p (L_0(p) + \omega) U_p^{-1}\,,
\end{equation}
with domain~$H^1(\R,\Cbb^2)$. A direct computation shows that
\begin{equation}\label{Formula_Ap}
	A_p = i p \sigma_2 \partial_x + m\sigma_3 - (p+1) g \sigma_3\,,
\end{equation}
where $g:\R\to(0,m-\omega]$ is given by
\[
	g(x) := (p+1)^{-1} \big|\tilde{v}_p^2(x)-\tilde{u}_p^2(x)\big|^p\,,
\]
and where $\tilde\phi_{0,p}=(\tilde v_p,\tilde u_p)\transp := p^{1/2}U_p\phi_{0,p}$, with $\phi_{0,p}=(v_p,u_p)\transp$ denoting the solitary wave.

At first sight, the function $g$ depends on~$p$. However, the key observation is that it is in fact \emph{independent} of~$p$, and is explicitly given by
\begin{equation}\label{Formula_g}
	g(x) = (m-\omega)\frac{1-\tanh^2(\kappa x)}{1-\nu\tanh^2(\kappa x)} = \frac{v_1^2 (x) - u_1^2 (x)}{2}\,,
\end{equation}
where $\kappa := \sqrt{m^2-\omega^2}$ and $\nu := (m-\omega)/(m+\omega)$. In particular, $g$ is entirely determined by the solitary wave corresponding to the Gross--Neveu case $p=1$. This follows from the explicit expressions of~$\phi_{0,p}$; see, for instance,~\cite{ChuPel-06, CooKhaMihSax-10,LeeKuoGav-75,MerQuiCooKhaSax-12,AldRicStoVDB-23}.

As a consequence, the family $\{A_p\}_{p>0}$ satisfies the identity
\begin{equation}\label{Identity_between_As}
	A_p = \frac{p}{q} A_q - \frac{p-q}{q} W\sigma_3\,, \qquad \forall\, p,q>0\,,
\end{equation}
where
\begin{equation}\label{Def_W}
	W := m - g\,.
\end{equation}
The function $W$ is even, increasing on~$[0,+\infty)$, and satisfies $W(0)=\omega < m = \|W\|_\infty$.

\medskip

The proof of Theorem~\ref{main_thm_1_gap_p_above_1}, for~$p\geq 1$, relies on the analysis of the family $\{A_p\}_{p>0}$ through the identity~\eqref{Identity_between_As}.
The key result allowing us to propagate spectral properties in~$p$ is the following proposition shown in Section~\ref{Section_proof_main_propagation_prop_global}.
\begin{proposition}\label{main_propagation_prop_global}
	Let $\omega\in(0,m)$ and~$p \geq q > 0$. Then,
	\[
		\sigma(A_q) \cap (-m, m) = \{-\omega, \omega\} \quad \Rightarrow \quad \sigma(A_p) \cap (-m, m) = \{-\omega, \omega\}\,.
	\]
\end{proposition}
Hence the absence of eigenvalues inside the spectral gap can be propagated along the parameter $p$.
At $q=1$, the absence of eigenvalues of~$A_1$ other than $\{\pm \omega\}$ is known from~\cite[Lemma~6.2]{AldRicStoVDB-23}. Proposition~\ref{main_propagation_prop_global} therefore extends this property to all $p> 1$.
This result is based on suitable applications of the min--max principle for operators with a spectral gap (see Paragraph~\ref{minmax} below).
 
 The second part of Theorem~\ref{main_thm_1_gap}, for~$p\in (0,1)$,
 follows 
 from Proposition~\ref{main_propagation_prop_global} combined
 with the
  following result applied to $q=1$, for which $A_1$ is known
 to admit threshold resonances.

\begin{proposition}\label{prop-birth}
	Let $(q,\omega)\in(0,+\infty)\times(0,m)$. If $A_q$ admits a resonance at $+m$, %then for~$\varepsilon>0$ small enough there exists an eigenvalue $\lambda_\varepsilon \in (\omega, m)$ of~$A_{q(1-\varepsilon)}$.
	then for $p<q$ sufficiently close to $q$ there exists an eigenvalue $\lambda_p \in (\omega, m)$ of~$A_p$.
	
	In particular, there exists $E_\star>0$ such that  for any $s\in(0,1)$, 
	\[
		m - 2m \frac{q-p}{q} \leq \lambda_p \leq m - E_\star \left(\frac{q-p}{q}\right)^{2+s} + o\!\left(\left(\frac{q-p}{q}\right)^{2+s}\right)\,.
	\]
	By symmetry of~$A_p$, we have $\pm\lambda_p \in \sigma(A_p) \cap (-m, m)$.
%	
%		
%		
%			Moreover, as $p \nearrow 1$, these eigenvalues satisfy
%				\[
%					m - 2m (1-p) \leq \pm (\lambda_\pm(p) + \omega) \leq m - (1-p)^{2+s} E_\star + o\!\left((1-p)^{2+s}\right)\,,
%				\]
%				for any $s\in (0,1)$ and where $E_\star>0$ is independent of~$p$.
\end{proposition}
\begin{remark*}\leavevmode
	\begin{enumerate}[label=\roman*)]
		\item The \emph{threshold energy drop} $E_\star$ is explicit and reads $\norm{\mathcal{E}}_1/ \norm{\psi^\infty}_\infty^2$, where $\mathcal{E}: \R \to \R$ defined below in~\eqref{Def_threshold_energy_density} is the \emph{threshold energy density} and $\psi^\infty$ the threshold resonance.
		\item By adapting the method of Cuenin and Siegl~\cite{CueSie-18}, one can obtain the the energy drop is of  order $(q-p)^2$. This method, however, requires very accurate knowledge of the resolvent of~$L_0$ at $q=1$ for energies close to the thresholds.
	\end{enumerate}
\end{remark*}
Since threshold resonances are known to exist in the Gross--Neveu case $q=1$, the two preceding propositions yield Theorem~\ref{main_thm_1_gap}.

Remarkably, they also imply the absence of threshold resonances for all $p>1$, as stated in Theorem~\ref{main_thm_noRes}. Indeed, if $A_q$ possessed a threshold resonance for some $q>1$, then there would be, by Proposition~\ref{prop-birth}, an eigenvalue inside the spectral gap of~$A_p$ for $p \in[1, q)$ close to $q$, contradicting Proposition~\ref{main_propagation_prop_global} (or Theorem~\ref{main_thm_1_gap_p_above_1}).

Moreover, Proposition~\ref{prop-birth} describes the mechanism of giving birth,
for $p<1$, to additional eigenvalues inside the spectral gap of~$L_0(p)$,
besides the distinguished eigenvalues $-2\omega$ and $0$. Thus, the
Gross--Neveu exponent $p=1$ separates two spectral regimes: the open gap
property propagates towards larger exponents, whereas the threshold
resonances at $p=1$ give birth to additional gap eigenvalues towards
smaller exponents.

The proof of Proposition~\ref{prop-birth}, given in
Section~\ref{Section_proof_prop_p_less_than_1}, combines the min--max
principle with a localization procedure for threshold resonances of~$A_q$.
The key point is that, after localizing a resonant state with a parameter
$\delta>0$, the scaling identity~\eqref{Identity_between_As} produces a
strictly negative contribution of order~$\varepsilon\delta$, while the
remaining terms are of higher order for a suitable choice of
$\delta=\delta(\varepsilon)$. This creates an energy drop below the
threshold~$m$, yielding an eigenvalue inside the spectral gap for
$\varepsilon>0$ sufficiently small.

\subsubsection{Min-max principle}
\label{minmax}
In this work we make extensive use of the \emph{min-max principle} (see,~e.g.,~\cite{DolEstSer-00,DolEstSer-00Errata,GriSie-99,DolEstSer-06,EstLewSer-19,SchSolTok-19}) for the family of operators $\{A_p\}_{p>0}$, defined in~\eqref{Def_Ap}, whose essential spectrum has a gap $(-m,m)$. More precisely, we use its version for self-adjoint operators; see, e.g.,~\cite{DolEstSer-00Errata} (for a version for symmetric operators, see, e.g.,~\cite{SchSolTok-19}).

For the application of this principle, one needs to construct orthogonal projections~$\Lambda_\pm$ on~$\mathcal{H}=L^2(\R, \Cbb^2)$, with the associated projection sets $F_\pm := \Lambda_\pm D(A_p)$ and the associated numbers
\[
	\gamma_k := \inf_{\substack{V \subset F_+ \\ \dim V = k}} \sup_{\varphi \in (V \oplus F_-) \setminus \{0\}} \frac{\langle \varphi, A_p \,\varphi \rangle_{ }}{ \norm{\varphi}_2^2}\,, \quad k \geq 0\,,
\]
satisfying the following conditions:
\begin{enumerate}
	\item {\it orthogonal decomposition:}
		\[
			\Lambda_+\mathcal{H} \oplus \Lambda_-\mathcal{H} = \mathcal{H} \quad \text{with} \quad F_\pm  \subset H^{1}(\R,\Cbb^2) \,;
		\]
	
	\item {\it gap condition:}
		\[
			\gamma_0 < \gamma_1\,.
		\]
\end{enumerate}
The min-max principle then states that
\[
	\gamma_\infty := \inf(\sigma_{\textrm{ess}}(A_p)\cap(\gamma_0,+\infty)) > \gamma_0
\]
and that the numbers $\gamma_k$ lying in~$(\gamma_0, \gamma_\infty)$ are exactly all the eigenvalues of~$A_p$ in the set $(\gamma_0, \gamma_\infty) $, counted with multiplicities. Moreover, if $\gamma_k \not\in[\gamma_0, \gamma_\infty)$, then $\gamma_k = \gamma_\infty$.
% If the multiplicity of~$\gamma_k$ is not finite, then it lies in the essential spectrum of~$A_p$.

We note that the $\gamma_k$'s are by definition ordered, $\gamma_0 \leq \gamma_1 \leq \gamma_2 \leq \dots$, and that the gap condition can equivalently be written as
\begin{align}\label{gap-con}
	\gamma_0 = \sup_{\varphi \in F_- \setminus \{0\}} \frac{\langle \varphi, A_p \varphi \rangle}{ \norm{\varphi}_2^2} < \inf_{\varphi_+ \in F_+\setminus\{0\}} \sup_{\varphi \in (\vect\{\varphi_+\} \oplus F_-) \setminus \{0\}} \frac{\langle \varphi, A_p \varphi \rangle}{ \norm{\varphi}_2^2} = \gamma_1\,.
\end{align}

\subsection{Generalized eigenfunctions at the thresholds}
In Section~\ref{Section_proof_main_thm_2_noEV}, we establish several structural properties of generalized eigenfunctions for one-dimensional Dirac operators of the form
\[
D_m+V\,,
\]
with particular emphasis on the threshold energies $\pm m$. In particular, we show that generalized eigenfunctions at the thresholds possess a rigid asymptotic structure: one component always belongs to $H^1(\R)$ and therefore vanishes at infinity, while the other component converges to constants and describes the possible resonance behavior.

The results obtained in Section~\ref{Section_proof_main_thm_2_noEV} provide several key ingredients for our analysis. First, they yield the qualitative information needed in the variational construction underlying Proposition~\ref{prop-birth}. Second, they provide the regularity and asymptotic properties required for the oscillation argument excluding threshold eigenvalues of~$L_0$, proving Theorem~\ref{main_thm_2_noEV}. Finally, their structure is also used to establish the simplicity of generalized eigenfunctions at the thresholds, leading to Theorem~\ref{main_thm_3_Resonances}.

Our results on simplicity can be formulated in a more general framework. In fact, Theorem~\ref{main_thm_3_Resonances} appears as a particular consequence of this more general statement.
\begin{theorem}[Simplicity of threshold generalized eigenvalues]\label{Thm_at_most_one_generalized_eigenfunction_at_thresholds}\leavevmode\nopagebreak
	Let $V \in L^1 \cap L^2 (\R,\Cbb^{2\times2})$ be such that $V-V\transp$ is purely imaginary-valued. If $\lambda=\pm m$, then 
	\[
	\dim \left\{ \text{generalized eigenfunctions of~$D_m + V$ associated to $\lambda$} \right\} \leq 1\,.
	\]
	
	In particular, this holds for~$D_m + V = L_\mu$, $\mu\in \R$.
\end{theorem}
Theorem~\ref{Thm_at_most_one_generalized_eigenfunction_at_thresholds} is proved in two steps. First, we establish a conditional simplicity result (Proposition~\ref{Proposition_at_most_one_generalized_eigenfunction_at_thresholds_conditional}) for operators $D_m+V$ with
\[
V \in L^1_{\rm loc}(\R,\Cbb^{2\times2})
\]
and $V-V\transp$ purely imaginary-valued. This criterion shows that a generalized eigenvalue is simple provided that all associated generalized eigenfunctions share a common component vanishing at the same point.

Second, Proposition~\ref{Proposition_Property_resonances_at_thresholds} ensures that this condition is satisfied at the thresholds when
\[
V\in L^1\cap L^2(\R,\Cbb^{2\times2})\,.
\]
Combining both results yields Theorem~\ref{Thm_at_most_one_generalized_eigenfunction_at_thresholds}.
\subsubsection{Parity}
For parity-preserving Dirac operators, the preceding structural results
can be further refined within each parity sector; see
Corollary~\ref{Corollary_simplicity_generalized_eigenfunction_under_parity} below. Since the operators $L_\mu$ belong to this
class, we introduce here the notation that will be used for this purpose.
We follow the convention used, e.g.,
in~\cite{BerCom-12,AldRicStoVDB-23}.

\begin{definition}[Even and odd subspaces]\label{Def_even_odd_function_subspaces}
	We say that a function $f: \R \mapsto \Cbb^2$ is \emph{even} (respectively \emph{odd}), if its first component is an even function (resp.~odd function) and its second component is odd (resp.~even). We denote by
	\[
	\even(\R, \Cbb^2) := \{ (f_1,f_2): \R \mapsto \Cbb^2 \ | \ (f_1,f_2)(-x) = (f_1,-f_2)(x) \}
	\]
	and
	\[
	\odd(\R, \Cbb^2) := \{ (f_1,f_2): \R \mapsto \Cbb^2 \ | \ (f_1,f_2)(-x) = (-f_1,f_2)(x) \}
	\]
	the corresponding subspaces.
	
	Moreover, an operator is said to be \emph{parity-preserving} if
	$\even(\R,\Cbb^2)$ and $\odd(\R,\Cbb^2)$ are invariant under its action.
	
	Finally, for~$p\in[1,+\infty]$, we define
	\[
	L^p_+(\R, \Cbb^2) := L^p(\R, \Cbb^2) \cap \even(\R, \Cbb^2) \quad \text{ and } \quad L^p_-(\R, \Cbb^2) := L^p(\R, \Cbb^2) \cap \odd(\R, \Cbb^2)\,.
	\]
	%	
	%	Moreover, for~$p\in[1,+\infty]$, we define
	%	\[
	%		L^p_\pm(\R, \Cbb^2) := \left\{(\phi_1,\phi_2)\transp \in L^p(\R, \Cbb^2)\,:\, (\phi_1,\phi_2)(-x) = (\pm\phi_1,\mp\phi_2)(x)\right\}\,.
	%	\]
\end{definition}
\begin{remark*}\leavevmode
	\begin{enumerate}[label=\roman*)]
		\item The subspaces $\even(\R, \Cbb^2)$ and $\odd(\R, \Cbb^2)$ are respectively the eigenspaces associated to the eigenvalues $+1$ and $-1$ of the \emph{spinor parity operator} $\sigma_3 \mathcal{P}$, where $\mathcal{P} f(x) := f(-x)$. Note that $D_m$ is parity preserving.
		\item Since each operator $L_\mu$ is parity preserving and all its eigenvalues are simple~\cite[Lemma~2.3]{AldRicStoVDB-23}, every eigenfunction of~$L_\mu$ belongs either to $L^2_+(\R,\Cbb^2)$ or to $L^2_-(\R,\Cbb^2)$.
		\item Finally, recall that, due to the $\sigma_1$-symmetry of~$L_0$, if $\phi$ is an eigenfunction of~$L_0+\omega$ associated with $\lambda$ and $\phi\in L^2_\pm(\R,\Cbb^2)$, then $\sigma_1\phi$ is an eigenfunction associated with $-\lambda$ and satisfies $\sigma_1\phi\in L^2_\mp(\R,\Cbb^2)$.
	\end{enumerate}
\end{remark*}

\subsection{Absence of threshold eigenvalues}\label{ah}
Finally, in Section~\ref{section_absence_of_threshold_eigenvalues} we prove
Theorem~\ref{main_thm_2_noEV} following the approach developed
in~\cite[proof of Theorem~1.3]{AldRicStoVDB-23}. The main idea is to use the
fact that $(L_0+\omega)^2$ is unitarily equivalent to a Schrödinger-type
operator. To see this write
\begin{equation}\label{Def_calA}
	L_0 + \omega = i\sigma_2\partial_x+M\sigma_3 \,,
\end{equation}
where
\begin{equation}\label{Def_M}
	M(x):=m-f\!\left(\pscal{\phi_0(x)}{\sigma_3\phi_0(x)}_{\Cbb^2}\right) = m-f(v^2(x)-u^2(x))
\end{equation}
on $\R$. Recalling~\eqref{Def_Ap}, we note that $(L_0 +\omega)$ is related to $A_p$ through
\[
\Par{L_0 +\omega}(p)= U_p^{-1} A_p U_p \,.
\]
Using the change of basis
\[
U=\frac{1}{\sqrt{2}}
\begin{pmatrix}
	1 & 1 \\*
	1 & -1
\end{pmatrix}
=\frac{1}{\sqrt{2}}(\sigma_1+\sigma_3)\,,
\]
we check that $\Par{L_0 +\omega}^2$ is unitarily equivalent to the block-diagonal operator
\begin{equation}\label{oldcA}
	U \Par{L_0 +\omega}^2 U =
	\begin{pmatrix}
		-\partial_x^2+M^2-M' & 0 \\
		0 & -\partial_x^2+M^2+M'
	\end{pmatrix},
\end{equation}
whose diagonal entries are Schrödinger operators with essential spectrum
$[m^2,+\infty)$. We then conclude the proof of
Theorem~\ref{main_thm_2_noEV} using Sturm--Liouville oscillation arguments.

In summary, squaring \(L_0+\omega\) reduces the threshold problem to two
scalar Schrödinger equations. A hypothetical threshold eigenfunction would
force the corresponding scalar solutions to have infinitely many zeros,
whereas a Sturm--Liouville comparison argument shows that threshold solutions
can have only finitely many zeros.

\section{Properties of generalized eigenfunctions at the thresholds}\label{Section_proof_main_thm_2_noEV}

In this section we establish several structural properties of generalized eigenfunctions of one-dimensional Dirac operators at the thresholds of the essential spectrum. These results constitute a key technical ingredient in our analysis. In particular, we show that generalized eigenfunctions at the thresholds possess a rigid asymptotic structure: one component always belongs to $H^1(\R)$, while the other converges to constants at infinity and describes the possible resonance behavior.

We begin with a general qualitative property of generalized eigenfunctions for one-dimensional Dirac operators. 
\begin{proposition}[Generic regularity of generalized eigenfunctions]\label{bas1}
	Let
	\[
		V \in L^1_{\rm loc}(\R,\Cbb^{2\times2})
	\]
	and $\Psi=(\psi_1,\psi_2)\transp \in L^\infty(\R,\Cbb^2)$ be a generalized eigenfunction of~$D_m+V$ associated with $\lambda\in\R$. Then, $\Psi\in AC(\R,\Cbb^2)$ and $\Psi(x)\neq 0$ for all $x\in\R$.
\end{proposition}
\begin{remark*}
	 Since $\Psi$ is continuous on~$\R$, a straightforward iteration in~$k\in \N\cup\{0\}$ of the eigenvalue equation implies that, if $V \in C^k(\R,\Cbb^{2\times2})$, then $\Psi\in C^{k+1}(\R,\Cbb^2)$.
\end{remark*}
We now turn to the special structure of generalized eigenfunctions at the threshold energies $\lambda=\pm m$. 
	\begin{proposition}[Properties of generalized threshold eigenfunctions] \label{Proposition_Property_resonances_at_thresholds}
		Let
		\[
			V\in L^1(\R,\Cbb^{2\times2}) \cap L^2(\R,\Cbb^{2\times2})
		\]
		and let $\Psi=(\psi_1,\psi_2)\transp \in L^\infty(\R,\Cbb^2)$ be a generalized eigenfunction of~$D_m+V$ at the threshold energy $+m$.
		Then,
		\begin{equation}\label{psi2inH1}
			\psi_2\in L^2(\R)\,, \qquad \psi_1'\in L^2(\R)\,, \qquad \text{and} \qquad \psi_2'\in L^1(\R)\cap L^2(\R)\,.
		\end{equation}
		
		Assume furthermore there exists $\varepsilon>0$ such that
		\[
			|V(x)| = \bigO(|x|^{-3-\varepsilon}) \qquad\text{as }|x|\to\infty\,.
		\]
		Then, $\psi_2\in L^1(\R)$ and there exist $l_\pm\in\R$ such that
		\begin{equation}\label{Resonances_form_of_nonL2_component}
			\psi_1 = l_+\chi_{(0,+\infty)} + l_-\chi_{(-\infty,0)} + \tilde\psi_1\,,
		\end{equation}
		with
		\[
			|\tilde\psi_1(x)|	= \bigO(|x|^{-\varepsilon/2}) \qquad\text{as }|x|\to\infty\,.
		\]
		
		Assume finally there exists $\varepsilon>0$ such that
		\[
			|V(x)| = \bigO(|x|^{-4-\varepsilon}) \qquad\text{as }|x|\to\infty\,.
		\]
		If $l_+=l_-=0$, then $\psi_1\in L^2(\R)$ and $\Psi$ is an eigenfunction.
\end{proposition}
\begin{remark}\leavevmode
	\begin{enumerate}[label=\roman*)]
		\item If $\lambda=+m$ then $\psi_1$ does not belong to $L^2(\R)$ in general. In fact, $\Psi$ is an eigenfunction precisely when $\psi_1\in L^2(\R)$, and it is a resonance otherwise with equation~\eqref{Resonances_form_of_nonL2_component} describing its structure. The key point for us is that at $\lambda=+m$ the second component always satisfies $\psi_2\in H^1(\R)$, hence $\psi_2(\pm\infty)=0$\,.
		
		\item If $|V(x)|=\bigO(e^{-2C|x|})$ as $|x|\to\infty$ for some $C>0$, then the same arguments as in our proof show that $|\psi_2(x)|=\bigO(e^{-C|x|})$ and $|\tilde\psi_1(x)|=\bigO(e^{-C|x|})$ in~\eqref{Resonances_form_of_nonL2_component}\,.

		\item The corresponding statement at the opposite threshold $\lambda=-m$ holds with the roles of~$\psi_1$ and $\psi_2$ exchanged. Indeed, applying $\sigma_1$ to the eigenvalue equation shows that the equation for~$V$ at $+m$ with $\Psi=(\psi_1,\psi_2)$ is equivalent to the equation for~$-\sigma_1 V\sigma_1$ at $-m$ with $\sigma_1\Psi=(\psi_2,\psi_1)$\,.
	\end{enumerate}
\end{remark}

\begin{remark*} 
 Note that property~\eqref{Resonances_form_of_nonL2_component} slightly generalizes results in~\cite{ErdGre-21}, see e.g. Lemma~5.1 and Lemma~5.2. Indeed, we show that property~\eqref{Resonances_form_of_nonL2_component} is satisfied by generalized eigenfunctions at the thresholds whenever $V \in L^1 \cap L^2(\R)$ with $|V| = \bigO(|\cdot|^{-3-\varepsilon})$ as $|x| \to \infty$, whereas in~\cite{ErdGre-21} this property is assumed to hold for potentials satisfying~$|V| \lesssim \langle \cdot \rangle^{-3-\varepsilon}$, where $\langle x\rangle:=\sqrt{1+|x|^2}$.
\end{remark*}

We now present the proofs of the two previous propositions.
\begin{proof}[Proof of Proposition~\ref{bas1}]
	Let $\Psi=(\psi_1,\psi_2) \in L^\infty(\R,\Cbb^2)$ be a generalized eigenfunction of~$D_m +V$ associated with $\lambda\in\R$, where $V \in L^1_{\rm loc}(\R,\Cbb^{2\times2})$. The generalized eigenvalue equation reads
	\begin{equation}\label{eveq}
		\Psi' = i\sigma_2 (m \sigma_3 - \lambda I_2) \Psi + i\sigma_2 V \Psi \,.
	\end{equation}
	in the distributional sense 	and, for any $K\subset \R$ compact, the distributional derivative~$\Psi'$ satisfies
	\[
		\norm{\Psi'}_{L^1(K)} \leq \max\{|m-\lambda|, |m+\lambda|\} |K| \norm{\Psi}_\infty + \norm{V}_{L^1(K)} \norm{\Psi}_\infty < +\infty \,.
	\]
	Thus, $\Psi \in \ac(K)$ on any such $K$ and therefore $\Psi \in \ac(\R)$. To prove that $\Psi$ does not vanish on~$\R$, suppose by contradiction that there exists $x_0 \in \R$ such that $\Psi(x_0)=0$. Then, $\Psi$ solves the Cauchy problem
\begin{equation}\label{cp1}
\left\{
\begin{aligned}
	 &\Psi' = i \sigma_2 (m\sigma_3 -\lambda \Id)\Psi + i\sigma_2 V \Psi\,,\\
	&\Psi(x_0) = 0\,,
\end{aligned}
\right.
\end{equation}
and we claim that $\Psi \equiv 0$, contradicting that $\Psi$ is a generalized eigenfunction, as in Definition~\ref{Def_generalized_eigenfunction}. Indeed, $\Psi \equiv 0$ because the equation of the Cauchy problem is a linear first-order system with $L^1_{\rm loc}$-coefficients and Carathéodory's theorem therefore guarantees the (existence and) uniqueness of an absolutely continuous solution to this problem; see, for example,~\cite[Theorem~5.3]{Hale-80} or \cite[supplements of §10 and §16]{Walter-98}.
\end{proof}
	
\begin{proof}[Proof of Proposition~\ref{Proposition_Property_resonances_at_thresholds}]

Assume now that $V \in L^1(\R) \cap L^2 (\R)$ . We write the eigenvalue equation for~$\lambda = m$ as
	\begin{equation}\label{EV_equation_at_m}
		\Psi' = im \sigma_2 (\sigma_3 - I_2) \Psi + i\sigma_2 V \Psi
		\Leftrightarrow
		\left\{
		\begin{aligned}
			\psi_1' &= (i\sigma_2 V \Psi)_1 - 2m \psi_2 \,,\\
			\psi_2' &= (i\sigma_2 V \Psi)_2 \,.
		\end{aligned}
		\right.
	\end{equation}
	We first observe that $\psi_2' \in L^1(\R)\cap L^2 (\R)$ since, for any $p\in [1,2]$,
	\begin{equation}\label{psi2inlp}
		\norm{\psi_2'}_p = \norm{(i\sigma_2 V \Psi)_2}_p \leq \norm{i\sigma_2 V \Psi}_p = \norm{V \Psi}_p \leq \norm{\Psi}_\infty \norm{V}_{p} < +\infty \,.
	\end{equation}
	Let
	\[
		D(x) := \overline{\psi_1}(x)\psi_2(x) - \overline{\psi_1}(-x)\psi_2(-x)\,.
	\]
	Using~\eqref{EV_equation_at_m} twice and integrating by parts, we obtain
	\begin{align*}
		\int_{-x}^{x} \overline{\psi_1} (i\sigma_2 V \Psi)_2 = \int_{-x}^{x} \overline{\psi_1} \psi_2' &= D(x) - \int_{-x}^{x} \overline{\psi_1}' \psi_2 \\
			&= D(x) + 2m \int_{-x}^{x} |\psi_2|^2 - \int_{-x}^{x} \overline{(i\sigma_2 V \Psi)_1}\, \psi_2 \,.
	\end{align*}
	Consequently, for any $x \geq 0$,
	\begin{align*}
		2m \int_{-x}^{x} |\psi_2|^2 &\leq |D(x)| + \int_{-x}^{x} |\psi_1|\, |(V \Psi)_2| + \int_{-x}^{x} |(V \Psi)_1|\, |\psi_2| \\
			&\leq 2 \norm{\Psi}_\infty^2 + \int_{-x}^{x} |\psi_1|\, |V \Psi| + \int_{-x}^{x} |V \Psi|\, |\psi_2| \\
			&\leq 2 \norm{\Psi}_\infty^2 + (\norm{\psi_1}_\infty + \norm{\psi_2}_\infty)\norm{\Psi}_\infty \int_{-x}^{x} |V| \leq 2 \norm{\Psi}_\infty^2 (1 + \norm{V}_1)\,.
	\end{align*}
	Therefore, $\psi_2 \in L^2(\R)$ and, in view of~\eqref{psi2inlp}, we conclude that $\psi_2 \in H^1(\R)$.
		
	Furthermore, using the first equation of the r.h.s.~of~\eqref{EV_equation_at_m}, we obtain that $\psi_1' \in L^2(\R)$ with
	\[
		\norm{\psi_1'}_{2} \leq \norm{V \Psi }_2 + 2m \norm{\psi_2}_{2} \leq \norm{\Psi}_{\infty} \norm{V}_{2} + 2m \norm{\psi_2}_{2}\,.
	\]
	
	\bigskip

	We turn to the proofs of the additional claims. 	
%	 In the following, we repeatedly use that
%	\[
%		\langle \cdot \rangle^{-1} \leq \min\{1,|\cdot|^{-1}\} \leq \sqrt{2}\,\langle \cdot \rangle^{-1} \quad \text{on } \R \,.
%	\]
%	
	\smallskip
	For the decay of~$\psi_2$, we observe that $\psi_2\in H^1(\R)$ implies that $\lim_{\pm\infty}\psi_2 = 0$ and that $\overline{\psi_2}\psi_2'$ is integrable on $\R$. Moreover, since $|V(x)|=\bigO(|x|^{-3-\varepsilon})$ as $|x|\to\infty$, there exists $T>0$ such that $|V|\lesssim |\cdot|^{-3-\varepsilon}$ on~$(-\infty, -T)$. The Lebesgue differentiation theorem therefore yields
	\begin{equation}\label{decayofpsi2}
		\begin{aligned}[b]
	 		|\psi_2(x)|^2 &= \left| \int_{-\infty}^{x} (|\psi_2|^2)' \right| = 2\left| \int_{-\infty}^{x} \Re(\overline{\psi_2}\psi_2') \right| = 2\left| \int_{-\infty}^{x} \Re\big(\overline{\psi_2}(i\sigma_2 V \Psi)_2\big) \right| \\ 
	 		&\leq 2 \norm{\psi_2}_\infty \norm{\Psi}_\infty \int_{-\infty}^{x} |V| \lesssim \int_{-\infty}^{x} |\cdot|^{-3-\varepsilon} = \frac{ |x|^{-2-\varepsilon} }{2+\varepsilon}\,, \qquad \forall\, x<-T \,.
			\end{aligned}
	\end{equation}
The case $x > T$ follows by integrating over $(x,\infty)$ and arguing analogously. Consequently, $\psi_2 \in L^1(\R)$, since $\psi_2 \in L^\infty(\R)$ and is therefore also integrable on~$[-T,T]$.
		
	\smallskip
		
	Under the assumption $|V(x)|=\bigO( |x |^{-3-\varepsilon})$ as $|x|\to \infty$, we present the proof of the convergence of~$\psi_1$ stated in~\eqref{Resonances_form_of_nonL2_component} and its rate only for~$x\to+\infty$; the case $x\to-\infty$ is analogous. First, since $\psi_2 \in L^1(\R)$, as established above, the first equation in~\eqref{EV_equation_at_m} implies $\psi_1' \in L^1(\R)$ since
	\[
		|\psi_1'| = |(i\sigma_2 V \Psi)_1 - 2m \psi_2| \leq \norm{\Psi}_\infty |V| + 2m |\psi_2| \in L^1(\R)\,.
	\]
	The Lebesgue differentiation theorem therefore gives
	\begin{equation}\label{Subsection_resonances_thresholds_property_key_inequality_cvgce_psi1}
		\begin{aligned}[b]
			|\psi_1(x) - \psi_1(y)| &= \left| \int_{y}^{x} \psi_1' \right| \leq \norm{\Psi}_\infty \int_{y}^{x} |V| + 2m \int_{y}^{x} |\psi_2| \\
				&\lesssim \int_{y}^{x} | \cdot |^{-3-\varepsilon} + | \cdot |^{-1-\frac{\varepsilon}{2}} \lesssim \int_{y}^{x} |\cdot|^{-1-\frac{\varepsilon}{2}} = \frac{y^{-\frac{\varepsilon}{2}} - x^{-\frac{\varepsilon}{2}}}{\frac{\varepsilon}{2}} \leq \frac{y^{-\frac{\varepsilon}{2}}}{\frac{\varepsilon}{2}}\,,
		\end{aligned}
	\end{equation}
	 for~$x,y>T$, where we assume $x>y$ without loss of generality.
	
	This estimate shows that $\{\psi_1(x_n)\}_n$ is a Cauchy sequence for any sequence $\{x_n\}_n$ diverging to $+\infty$, and therefore converges.
	
	Now consider two sequences $\{x_n\}_n$ and $\{y_n\}_n$ diverging to $+\infty$. Both $\{\psi_1(x_n)\}_n$ and $\{\psi_1(y_n)\}_n$ are convergent; denote by $l_+$ the limit of~$\{\psi_1(x_n)\}_n$. Let $\eta>0$. Using~\eqref{Subsection_resonances_thresholds_property_key_inequality_cvgce_psi1}, choose $M$ such that $|\psi_1(x)-\psi_1(y)|\leq\eta$ for all $x,y\geq M$. By the divergence of~$\{x_n\}_n$ and the convergence of~$\{\psi_1(x_n)\}_n$, choose $N$ such that $x_N\geq M$ and $|\psi_1(x_N)-l_+|\leq\eta$. By the divergence of~$\{y_n\}_n$, choose $N'$ such that $y_n\geq M$ for all $n\geq N'$. Then,
	\[
		|\psi_1(y_n)-l_+| \leq |\psi_1(y_n)-\psi_1(x_N)| + |\psi_1(x_N)-l_+| \leq 2\eta
	\]
	for all $n\geq N'$. Hence, $l_+$ is also the limit of~$\{\psi_1(y_n)\}_n$. This proves that $\psi_1(x)$ converges as $x\to+\infty$, since the limit along any sequence diverging to $+\infty$ is the same, establishing~\eqref{Resonances_form_of_nonL2_component}.
	
	For the proof of the rate of convergence of~$\tilde\psi_1$, we again only present the argument for~$x\to+\infty$, which follows by taking this limit in~\eqref{Subsection_resonances_thresholds_property_key_inequality_cvgce_psi1}:
	\[
		\forall\, y>T\,, \quad |\tilde\psi_1(y)| = |l_+ - \psi_1(y)| = \lim_{x\to+\infty} |\psi_1(x) - \psi_1(y)| \leq C_\varepsilon y^{-\frac{\varepsilon}{2}} \,.
	\]
	where $C_\varepsilon$ is a constant depending on~$\varepsilon$ but not on~$y$.
	
	\bigskip
	
	Finally, for the proof of this rate under the assumption $|V| = \bigO(|x|^{-4-\varepsilon})$, the analogous computations as in~\eqref{decayofpsi2} and in~\eqref{Subsection_resonances_thresholds_property_key_inequality_cvgce_psi1}, under this stronger assumption on $V$, yields
	\[
		|\psi_1(x) - \psi_1(y)| \leq \tilde{C}_\varepsilon y^{-\frac{1+\varepsilon}{2}}
	\]
	and consequently that $\tilde\psi_1\in L^2(\R)$ since
	\[
		\forall\, y>T\,, \quad |\tilde\psi_1(y)| \leq \tilde{C}_\varepsilon y^{-\frac{1+\varepsilon}{2}} \,.
	\]
	Therefore, if $l_+ = l_- = 0$ in~\eqref{Resonances_form_of_nonL2_component}, then $\psi_1 = \tilde\psi_1 \in L^2(\R)$ and $\Psi$ is an eigenfunction.
\end{proof}

\section{Propagation of the open gap property}\label{Section_proof_main_propagation_prop_global}
This section is devoted to the proof of Proposition~\ref{main_propagation_prop_global} and therefore of Theorem~\ref{main_thm_1_gap_p_above_1}. 
The result is obtained by iterating the following local propagation statement. 
Its proof relies on the min--max principle for operators with a spectral gap.
\begin{proposition}\label{main_propagation_prop_local}
	Let $(q,\omega)\in(0,+\infty)\times(0,m)$. If $p \geq q$ satisfies
	\begin{equation}\label{main_prop_range_p}
		\frac{p}{q} < \frac{m}{m-\omega}\,,
	\end{equation}
	then
	\[
		\sigma(A_q) \cap (-m, m) = \{-\omega, \omega\} \quad \Rightarrow \quad \sigma(A_p) \cap (-m, m) = \{-\omega, \omega\}\,.
	\]
\end{proposition}

We first show how Proposition~\ref{main_propagation_prop_local} implies Proposition~\ref{main_propagation_prop_global}, then we prove the former.
\begin{proof}[Proof of~Proposition~\ref{main_propagation_prop_global}]
	%Let $\omega\in(0,m)$ be fixed.
	An application of Proposition~\ref{main_propagation_prop_local} to~$q>0$, for which we assume $\sigma(A_q) \cap (-m, m) = \{-\omega, \omega\}$, yields the claimed property of~$\sigma(A_p)$ for~$p\in[q, q(\alpha-\varepsilon)]$ and for any $\varepsilon\in(0,\alpha-1)$ where
	\[
		\alpha := \min\left\{ \frac{m}{m-\omega}, \frac{2m}{m+\omega} \right\} > 1.
	\]
	We fix such~$\varepsilon>0$. A second application, this time to~$q(\alpha-\varepsilon)$, gives the claimed property for~$p\in[q(\alpha-\varepsilon), q(\alpha-\varepsilon)^2]$, hence for~$p\in[q, q(\alpha-\varepsilon)^2]$. Since $\alpha-\varepsilon>1$, applying Proposition~\ref{main_propagation_prop_local} successively to~$q(\alpha-\varepsilon)^n$, $n\in\N$, proves Proposition~\ref{main_propagation_prop_global}.
\end{proof}

In order to prove Proposition~\ref{main_propagation_prop_local}, we will repeatedly use the identity \eqref{Identity_between_As}
\[
	A_p = \frac{p}{q} A_q - \frac{p-q}{q} W\sigma_3\,, \qquad \forall\, p, q>0\,.
\]
Note that, by definition of~$g$, the operator $W\sigma_3$ satisfies
\begin{equation}\label{bounds_m_minus_g_sigma3}
	-m \leq W\sigma_3 \leq +m \qquad \text{on } L^2(\R, \Cbb^2)\,.
\end{equation}
With these considerations we are ready to prove the desired result.
\begin{proof}[Proof of Proposition~\ref{main_propagation_prop_local}]
	Assume that
	\[
		\sigma(A_q)\cap(-m, m)=\{-\omega,+\omega\}\,.
	\]
		
	In order to apply the min–max principle (see Subsection~\ref{minmax}), we first construct the admissible orthogonal decomposition. Let $E_I(A)$ denote the spectral projector of a self-adjoint operator $A$ associated with the set $I\subset\R$. We define the orthogonal projections~$\Lambda_\pm$ on the Hilbert space~$\cH=L^2(\R, \Cbb^2)$ as
	\[
		\Lambda_- := P_- + P_{-\omega} \quad \text{and} \quad \Lambda_+ := P_{+\omega} + P_+\,,
	\]
	with
	\[
		P_{\alpha} := E_{\{\alpha\}}(A_q)\,, \ \forall\, \alpha \in \R\,, \quad P_- := E_{(-\infty,-m]}(A_q)\,, \quad \text{and} \quad P_+ := E_{[+m,+\infty)}(A_q)\,,
	\]
	and set
	 \[
		 F_\pm := \Lambda_\pm D(A_q) = \Lambda_\pm H^1(\R,\Cbb^2)\,.
	 \]
	 As required, $\Lambda_+\mathcal{H} \oplus \Lambda_-\mathcal{H} = \mathcal{H}$, by assumption on~$\sigma(A_q)$, and $F_\pm \subset H^1(\R,\Cbb^2) = D(A_q)$. Moreover, we note that
	 \[
		 P_- = E_{(-\infty,-\omega)}(A_q) \qquad \text{and} \qquad P_+ = E_{(+\omega,+\infty)}(A_q).
	 \]
%	 We then define
%	 \[
%		 F_\pm := \Lambda_\pm H^1(\R,\Cbb^2)\,,
%	 \]
%	 which satisfies $F_\pm := \Lambda_\pm H^1(\R,\Cbb^2) \subset H^1(\R,\Cbb^2) = D(A_q).$, as required.
	
	Consider now $p\geq q$ satisfying~\eqref{main_prop_range_p}.
	First, we have the upper bound
	\begin{align*}
		\gamma_0 &= \sup\limits_{h\in F_-\setminus\{0\}} \frac{\pscal{h}{A_p h}}{ \norm{h}_2^2} = \sup\limits_{h\in F_-\setminus\{0\}} \left[ \frac{p}{q} \frac{ \pscal{h}{A_q h} }{ \norm{h}_2^2 } - \frac{p-q}{q} \frac{ \pscal{h}{W\sigma_3 h} }{ \norm{h}_2^2 } \right] \\
			&\leq \sup\limits_{h\in F_-\setminus\{0\}} \left[ \frac{p}{q} \frac{ \pscal{h}{A_q h} }{ \norm{h}_2^2 } + \frac{p-q}{q}m \right] = -\frac{p}{q} \omega + \frac{p-q}{q}m = -m + \frac{p}{q} (m-\omega) < 0\,,
	\end{align*}
	where the last inequality is assumption~\eqref{main_prop_range_p}. In particular, this ensures~$\gamma_\infty \leq m$.
	
	Second, we have the lower bound
	\begin{align*}
		\gamma_1 := &{} \inf\limits_{\substack{V\subset F_+\\\dim V = 1}}\sup\limits_{h\in (V\oplus F_-)\setminus\{0\}}\frac{\pscal{h}{A_p h}}{ \norm{h}_2^2} \\
		\geq&{} \inf\limits_{h\in F_+\setminus\{0\}} \frac{\pscal{h}{A_p h}}{ \norm{h}_2^2} = \inf\limits_{h\in F_+\setminus\{0\}} \left[ \frac{p}{q} \frac{ \pscal{h}{A_q h} }{ \norm{h}_2^2 } - \frac{p-q}{q} \frac{ \pscal{h}{W\sigma_3 h} }{ \norm{h}_2^2 } \right] \\
		\geq&{} \inf\limits_{h\in F_+\setminus\{0\}} \left[ \frac{p}{q} \frac{ \pscal{h}{A_q h} }{ \norm{h}_2^2 } - \frac{p-q}{q} m \right] = \frac{p}{q} \omega - \frac{p-q}{q} m = m - \frac{p}{q} (m-\omega) >0\,,
	\end{align*}
	where we used $0 \in F_-$ for the first inequality and where the final positivity is again assumption~\eqref{main_prop_range_p}. Therefore, $\gamma_1 > \gamma_0$ and the min-max principle applies.
	
	Third, by the definition of~$F_+$ and due to the simplicity of eigenvalues of~$L_0$ (see \cite[Lemma 2.3]{AldRicStoVDB-23}), we have the decomposition 
	\[
		F_+ = P_{+\omega} H^1 \oplus P_{+} H^1 = \vect\{\phi_{0,q}\} \oplus P_{+} H^1\, .
	\]
	Therefore, for any~$V\subset F_+$ with $\dim V \geq2$ there exists $h_0\in V\setminus\{0\}$ such that~$h_0\perp \phi_{0,q}$. This allows us to establish the lower bound
	\begin{align*}
		\gamma_2 := &{} \inf\limits_{\substack{V\subset F_+\\\dim V = 2}} \sup\limits_{h\in (V\oplus F_-)\setminus\{0\}} \frac{\pscal{h}{A_p h}}{ \norm{h}_2^2} \geq \inf\limits_{\substack{V\subset F_+\\\dim V = 2}} \sup\limits_{\substack{h_0 \in V\setminus\{0\}\\ h_0\perp \phi_{0,q}}} \frac{\pscal{h_0}{A_p h_0}}{ \norm{h_0}_2^2} \\
		\geq&{} \inf\limits_{\substack{V\subset F_+\\\dim V = 2}} \sup\limits_{\substack{h_0 \in V\setminus\{0\}\\ h_0\perp \phi_{0,q}}} \left[ \frac{p}{q} \frac{ \pscal{h_0}{A_q h_0} }{ \norm{h_0}_2^2 } - \frac{p-q}{q} m \right] = \frac{p}{q} m - \frac{p-q}{q} m = m\,.
	\end{align*}
	
	To summarize, \cite[Theorem~1]{DolEstSer-00Errata} applies, since the \emph{orthogonal decomposition} is satisfied and since we have established the \emph{gap condition} $\gamma_0 < \gamma_1$ for~$p$ in the range defined by~\eqref{main_prop_range_p}. It first guarantees that $\gamma_\infty \geq \gamma_2 \geq m$ and, consequently, that $\gamma_\infty = \gamma_2 = m$ since we established $\gamma_\infty \leq m$ earlier. Second, it ensures that the $\gamma_k$'s such that $\gamma_k \in (\gamma_0, \gamma_\infty) = (\gamma_0, m)$ are exactly all the eigenvalues of~$A_p$ in this interval. Since~$\gamma_2 = m$, we deduce that $(\gamma_0,m)$ contains at most one eigenvalue and, if it does, then its value is~$\gamma_1$. Since~$\omega \in (0, m) \subset (\gamma_0,m)$ is always an eigenvalue of~$A_p$, we deduce\footnote{Another consequence, which is worth noticing for the clarity of the argument even though we do not use it, is that $\gamma_0 \geq - \omega$. Indeed, the contrary would yield the contradiction that $(\gamma_0,m)$ contains at most one eigenvalue while the known eigenvalues $\pm\omega$ are in~$(\gamma_0,m)$.} that $\gamma_1 = \omega \in(0, m)$. Therefore,
	\[
		\sigma(A_p) \cap (\gamma_0, m) = \{\omega\},
	\]
	and, by the symmetry of the spectrum of~$A_p$ with respect to~$0$ and since $\gamma_0<0$, that
	\[
		\sigma(A_p) \cap (-m, m) = \{-\omega, +\omega\}. \qedhere
	\]
\end{proof}

\section{Threshold resonances and the birth of gap eigenvalues}\label{Section_proof_prop_p_less_than_1}
This section is devoted to the proof of Proposition~\ref{prop-birth} and therefore of Theorem~\ref{main_thm_1_gap_p_below_1}. Let $q > 0$. The starting point is to consider for $p<q$ sufficiently close to $q$, $A_p$ as a perturbation of~$A_q$ through the identity~\eqref{Identity_between_As}. Throughout this section we set
\[
	p := (1-\varepsilon)q < q \quad \Leftrightarrow \quad \varepsilon := \frac{q-p}{q} > 0\,.
\]
Hence, we have 
\begin{equation}\label{Identity_between_As_for_p_below_1}
	A_{p} = (1-\varepsilon)A_q +\varepsilon W \sigma_3
\end{equation}
and we recall that $\norm{W}=\norm{m-g}_\infty=m$. 

\medskip

The proof combines the scaling identity~\eqref{Identity_between_As} with the threshold resonance states of~$A_q$ and a suitable application of the min--max principle. More precisely, we construct trial states by localizing the resonance at the upper threshold $m$. The localization parameter $\delta>0$ regularizes the resonance and produces admissible $L^2$ trial states while preserving the leading threshold behavior. When inserted into the quadratic form of~$A_p$, these states generate an energy strictly below $m$.

The key point is that the main contribution, which will be estimated in Lemma~\ref{lem:first-term}, yields a negative term of order $\varepsilon\delta$, while all remaining contributions are of higher order for a suitable choice of~$\delta=\delta(\varepsilon)$; see in particular Lemma~\ref{lem-norm}, Corollary~\ref{cor:cross}, and~\eqref{UpperBound_gamma0}. This produces an energy drop below the threshold and therefore yields an eigenvalue in~$(\omega,m)$ for~$\varepsilon>0$ sufficiently small.

The proof is organized as follows. After two preliminary results, we introduce the spectral projections needed for the min--max principle and study the threshold resonance states of~$A_q$. We then construct the localized trial states and formulate the min--max strategy. The remainder of the section is devoted to the estimates required to prove $\gamma_1<m$ and to their combination with a suitable choice of~$\delta=\delta(\varepsilon)$.

\subsection{Preliminary results}

We begin with two lemmas that will be used in the proof, concerning the finiteness of the number of eigenvalues in the gap $(-m,m)$ and their regularity with respect to~$p>0$. We emphasize that both results are only needed for~$p\in(0,1)$: when $p\geq1$, Theorem~\ref{main_thm_1_gap_p_above_1} established in the previous section guarantees finiteness, since $A_p$ admits only the two eigenvalues $\pm\omega$ independently  of~$p$.

\begin{lemma}\label{Lemma_analyticity_eigenvalues}
 	Let $\omega \in (0,m)$. Any branch~$p \mapsto \lambda(p, \omega)$ of eigenvalues of $A_p$ inside the gap~$(-m, m)$ is analytic on the interval on which it exists.
\end{lemma}
The proof relies on Kato theory.
\begin{proof}
	Write $A_p = p \tilde{A}_{1/p}$ with
	\[
		\tilde{A}_r := i \sigma_2 \partial_x - g \sigma_3 + r W\sigma_3 \,.
	\]
	Proving that the eigenvalues of $\tilde{A}_r$ are analytic in the parameter $r>0$ yields the analyticity in $p>0$ of those of $A_p$.
	
	The family $\{\tilde{A}_r\}_{r>0}$ is linear (hence holomorphic) in $r$, with fixed domain $H^1(\R)$ and an $L^2(\R)$-bounded perturbation. It is therefore a \emph{holomorphic family of type (A)} in the sense of Kato, and the eigenvalues are consequently analytic in the parameter $r$ on the interval on which they respectively exist; see~\cite{Kato-50a,Kato-95-book}.
\end{proof}

To complete the picture regarding each individual eigenvalue, we recall that the eigenvalues inside the gap $(-m, m)$ are simple (see, e.g.,~\cite[Lemma~2.3]{AldRicStoVDB-23}) and are isolated as part of the discrete spectrum.

\smallskip

Note also that since we consider the whole range $q>0$ for our reference operator $A_q$, we cannot exclude the presence of eigenvalues of~$A_q$ in~$(\omega, m)$, as we could do for $q>1$; see Theorem~\ref{main_thm_1_gap_p_above_1} proved in the previous section. Nevertheless, due to the exponential decay of~$g$, 
%Lemma~\ref{Lemma_Schr_eigenfunction_at_threshold_vanish_finitely_many_times} allows us to deduce that
$A_q$ defined in~\eqref{Formula_Ap} is a massive Dirac operator with a potential that decays fast enough to ensure that the eigenvalues of the operator do not accumulate at the edges of its essential spectrum; this is the result presented in the next lemma. Therefore, there exists a largest eigenvalue of~$A_q$ in~$[\omega, m)$:
\[
	\exists\, \tau \in \sigma(A_q) \cap [\omega, m)\,, \quad \sigma(A_q) \cap (\tau, m) = \emptyset\,.
\]
 
\begin{lemma}\label{Lemma_thresholds_Aq_not_accumulation_points}
 	Let $(q,\omega)\in(0,+\infty)\times(0,m)$. Then, $A_q$ has finitely many eigenvalues in the spectral gap~$(-m, m)$. In particular, eigenvalues cannot accumulate at the thresholds~$\pm m$.
\end{lemma}
According to the discussion in Subsection~\ref{ah}, it is enough to
establish the finiteness of the number of negative eigenvalues of the
Schrödinger operators $-\partial_x^2+M^2\mp M'-m^2$. This finiteness
immediately implies Lemma~\ref{Lemma_thresholds_Aq_not_accumulation_points}.

\begin{proof}
	Using~\eqref{oldcA}, one can check from the explicit formulae for $M$
	and $M'$ that
	$$
	V_\pm:=M^2\mp M'-m^2
	$$
	decays exponentially. Therefore,
	$x\mapsto |x|\max\{0,-V_\pm(x)\}\in L^1(\R)$, and the finiteness of the
	number $N(V_\pm)$ of negative eigenvalues of
	$-\partial_x^2+V_\pm$ follows from the one-dimensional Bargmann-type
	inequality~\cite{ChaKhuMarWu-03} (going back
	to~\cite{Bargmann-52,Schwinger-61}; see
	also~\cite{MolVai-12,MolVai-23}):
	$$
	N(V_\pm)
	\leq
	1+\int_{\R}|x|\max\{0,-V_\pm(x)\}\,\di x
	<\infty\,.
	$$
	Since $U(L_0+\omega)^2U$ has only finitely many eigenvalues below
	$m^2$, the same holds for $(L_0+\omega)^2$. Consequently,
	$L_0+\omega$ has only finitely many eigenvalues in $(-m,m)$, and by
	unitary equivalence, so does $A_p$, concluding the proof.\qedhere
\end{proof}

Now we are ready to begin with the proof of Proposition~\ref{prop-birth}.

\subsection{Admissible projections.}\label{Subsect_Admissible_proj}
In order to apply the min-max principle and detect an eigenvalue above $\tau$, we introduce the orthogonal decomposition of~$\cH := L^2(\R,\Cbb^2)$ defined through
\begin{equation}\label{Def_Lambda_pm_NearBelow_1}
	\Lambda_- := E_{(-\infty,\tau]}(A_q)\quad \text{and} \quad \Lambda_+ := E_{[m,+\infty)}(A_q)\,.
\end{equation}
We then set
\[
	F_\pm := \Lambda_\pm H^{1}(\R,\Cbb^2) \subset H^{1}(\R,\Cbb^2) = D(A_q)\,.
\]
Since $\sigma(A_q) \cap (\tau,m) = \emptyset$ by definition of~$\tau$, the projections $\Lambda_\pm$ define an orthogonal decomposition of~$\cH$. Namely, $\Lambda_+\mathcal{H} \oplus \Lambda_-\mathcal{H} = \mathcal{H}$.

We now derive an upper bound on~$\gamma_0$ associated with the subspace $F_-$. Using~\eqref{Identity_between_As_for_p_below_1}, we obtain
\begin{equation}\label{UpperBound_gamma0}
	\pscal{\varphi}{A_p \varphi} \leq \big((1-\varepsilon)\tau + \varepsilon \norm{W}\big)\norm{\varphi}_2^2 \leq \big(\tau + \varepsilon (m-\tau)\big)\norm{\varphi}_2^2\,, \quad \forall\, \varphi \in F_-
\end{equation}
and consequently
\[
	\gamma_0 = \sup\limits_{h\in F_-\setminus\{0\}} \frac{\pscal{h}{A_p h}}{ \norm{h}_2^2}\leq \tau + \varepsilon (m-\tau)\,.
\]

%\begin{remark*}
%	We note that $\gamma_0 \geq \omega$. Indeed, since the solitary wave $\phi_0 (q)\equiv \phi_0$ satisfies $A_q\phi_0 = \omega \phi_0$, we have $\phi_0 \in F_-\setminus\{0\}$ and
%	\[
%		\gamma_0 \geq \frac{\pscal{\phi_0}{ A_p \phi_0}}{ \norm{\phi_0}_2^2} = (1-\varepsilon)\omega + \varepsilon \frac{\pscal{\phi_0}{ W \sigma_3 \phi_0}}{ \norm{\phi_0}_2^2} = \omega\,,
%	\]
% where $\pscal{\phi_0}{ W \sigma_3 \phi_0} = \omega \norm{\phi_0}_2^2$ by an explicit computation.
%\end{remark*}

In addition, recalling the formula of~$\gamma_1$ given in~\eqref{gap-con}, we have 
\[
	\gamma_1= \inf_{\varphi_+ \in F_+\setminus\{0\}} \sup_{\varphi \in (\vect\{\varphi_+\} \oplus F_-) \setminus \{0\}} \frac{\langle \varphi, A_p \varphi \rangle}{ \norm{\varphi}_2^2} \geq \inf_{\varphi \in F_+ \setminus \{0\}} \frac{\langle \varphi , A_p \varphi \rangle}{ \norm{\varphi}_2^2}\geq (1-2\varepsilon)m\,,
\]
where we choose $\varphi = \varphi_+$ for the lower bound on the $\sup$, then use~\eqref{bounds_m_minus_g_sigma3}.

Hence, the \emph{gap condition} $\gamma_0<\gamma_1$ in~\eqref{gap-con}, necessary to apply the min-max principle, is satisfied for any $\varepsilon<\varepsilon_0 := (m-\tau)/(3m-\tau)$.
Therefore, under this smallness condition on~$\varepsilon$, $\Lambda_\pm$ give an orthogonal decomposition of~$\cH$ suitable for the application of the min-max principle.

Moreover, we claim that $\gamma_\infty = m$ under this smallness condition. Indeed, the upper bound on $\gamma_0$ guarantees, through the definition of $\gamma_\infty$, that $\gamma_\infty \leq m$ if $\varepsilon \leq 1$, which is ensured by $\varepsilon<\varepsilon_0$. Meanwhile, the lower bound on $\gamma_1$ yields $\gamma_\infty \geq \gamma_1 \geq (1-2\varepsilon)m$, hence $\gamma_\infty > -m$, due to the smallness condition on $\varepsilon$, hence $\gamma_\infty \geq m$ due to its definition.

%\medskip
\subsection{Properties of resonance states.}

A key ingredient in our analysis is the use of the properties of possible threshold resonance states of~$A_q$. We use them to construct suitable trial states for the min-max principle.

More precisely, assume that $A_q$ admits a resonance at each threshold $\pm m$ of its essential spectrum. That is, there exist $\psi_\pm^{\infty}\in L^\infty(\R,\Cbb^2)\setminus L^2 (\R)$ solutions to
\begin{equation}\label{res}
	A_q\psi_\pm^\infty=\pm m \psi_\pm^\infty\,.
\end{equation}
By symmetry, we have $\psi_+^\infty=\sigma_1 \psi_-^{\infty}$.

\begin{remark*}
We do not consider the case in which the generalized eigenfunction at the threshold is actually an eigenfunction, since in Section~\ref{section_absence_of_threshold_eigenvalues} we will prove the absence of threshold eigenvalues. Moreover, constructing the aforementioned trial states from a threshold eigenvalue appears to be simpler. Indeed, the only place where we need to assume that $\psi^\infty \notin L^2(\R)$ is in Lemma~\ref{C_infty_positive} below.
\end{remark*}
	
 In what follows we focus on the threshold at $+m$ and write
 \[
 	\psi^\infty \equiv \psi_+^\infty =: (\psi_1^\infty, \psi_2^\infty)\transp\,.
 \]
Recall that $\psi^\infty$ is absolutely continuous on~$\R$ by Proposition~\ref{bas1}. Moreover, by Proposition~\ref{Proposition_Property_resonances_at_thresholds} and since $g$ decays exponentially fast, we have
\[
	\psi_2^\infty \in H^1(\R)\quad \text{while} \quad \psi_1^\infty\notin L^2
\]
since it converges to constant values as $x\to\pm \infty$, which are non-zero as $\psi^\infty$ would otherwise be an eigenfunction.
	
Our next result shows that $|\psi^\infty|$ is bounded away from zero. The proof crucially uses that $|\psi^\infty|$ is not an eigenfunction ($\psi_1^\infty$ is not square integrable), together with the parity symmetry. This result will be used later on to control the normalization of the trial states.
\begin{lemma}\label{C_infty_positive}
	Let $(q,\omega)\in(0,+\infty)\times(0,m)$. If $\psi^\infty$ is a threshold resonance of~$A_q$ at~$+m$, then
	\begin{equation}\label{def_c_infty}
		c_\infty := \inf_{\R} |\psi^\infty|^2 >0\, .
	\end{equation}
\end{lemma}
\begin{proof}
	From the resonance equation~\eqref{res} and the fact that $A_q$ preserves parity, we may assume that $\psi^\infty$ belongs to one of the parity subspaces $L^\infty_\pm$ introduced in Definition~\ref{Def_even_odd_function_subspaces}. Therefore, $|\psi^\infty|^2$ is an even function on~$\R$.
	
By Proposition~\ref{bas1}, we have 
\[
	|\psi^\infty|^2=|\psi_1^\infty|^2+|\psi_2^\infty|^2>0\,, \quad \text{on } \R\,.
\]
Moreover, $\psi_2^\infty \in H^1(\R)$ by Proposition~\ref{Proposition_Property_resonances_at_thresholds} and then $|\psi_2^\infty(x)| \to 0$ as $|x|\to\infty$.
 
Suppose now, by contradiction, that $c_\infty = 0$. Since $|\psi^\infty|^2>0$ on~$\R$ and $|\psi^\infty|^2$ is even, this implies $|\psi^\infty(\pm \infty)|=0$ and consequently $l_\pm =0$ (from Proposition~\ref{Proposition_Property_resonances_at_thresholds}). In particular, and still by Proposition~\ref{Proposition_Property_resonances_at_thresholds}, $\psi_1^\infty \in L^2(\R)$ and consequently $\psi^\infty \in L^2 (\R,\Cbb^2)$, contradicting that $\psi^\infty$ is a resonance (Definition~\ref{Def_generalized_eigenfunction}).
\end{proof}
\begin{remark}\label{rmres} Threshold resonances are known explicitly for the operator $A_1$; see, e.g. \cite[Lemma~6.1]{AldRicStoVDB-23}. Using the notation $\phi_0 = (v, u)\transp$ for the eigenfunction of~$A_1$ to the eigenvalue~$\omega$, we have
\begin{equation}\label{def-psi}
	\begin{aligned}
		&\psi_1^\infty = \frac{uv}{v^2-u^2} = \frac{ \sqrt{\nu} \tanh(\kappa \cdot) }{ 1 - \nu \tanh^2(\kappa \cdot) } \quad \text{and}\\
		&\psi_2^\infty = -\frac{\nu}{1-\nu} \frac{ v^2 - \nu^{-1} u^2 }{ v^2-u^2 } = -\frac{\nu}{1-\nu} \frac{ 1 - \tanh^2(\kappa \cdot) }{ 1 - \nu \tanh^2(\kappa \cdot) }\,.
	\end{aligned}
\end{equation}
which explicitly verifies 
\[
	0<c_\infty = \frac{1}{4 \omega^2}
	\left\{
	\begin{aligned}
		&\frac{m^2 - 2\omega^2}{2}\,, \quad &&\text{if } 0 < \omega \leq \frac{m}{2}\,, \\
		&(m-\omega)^2\,, \quad &&\text{if } \frac{m}{2} \leq \omega < m\,.
	\end{aligned}
	\right.
\]

\end{remark}
Using the resonance equation \eqref{res} for~$\psi^\infty$ together with the identity
\eqref{Identity_between_As_for_p_below_1}, we compute the pointwise energy density of~$\psi^\infty$ for the operator $A_p$ and obtain
\begin{equation}\label{obs}
	\pscal{\psi^\infty(x)}{A_p (x)\psi^\infty(x)}_{\Cbb^2} = m \left| \psi^\infty(x) \right|^2 - \varepsilon \mathcal{E}(x)\,,
\end{equation}
where $\mathcal{E}$ is the \emph{threshold energy density} defined by
\begin{equation}\label{Def_threshold_energy_density}
	\mathcal{E}(x) := m \left| \psi^\infty(x) \right|^2 - W(x) \left\langle \psi^\infty(x),\sigma_3\, \psi^\infty(x) \right\rangle_{\Cbb^2}\,.
\end{equation}
The key point is that
\[
	\mathcal{E}(x)>0\,,\quad \forall\, x\in \R\,.
\]
In particular, \eqref{obs} strongly suggests that the threshold resonance of~$A_q$ produces an energy drop for~$A_p$ only when $\varepsilon>0$, i.e., when $p<q$.
\begin{lemma}\label{lem:energy-density}
	Let $(q,\omega)\in(0,+\infty)\times(0,m)$, $\psi^\infty \equiv \psi_+^\infty =: (\psi_1^\infty, \psi_2^\infty)\transp \in L^\infty(\R,\Cbb^2)$ be a threshold resonance of~$A_q$ at $m$, and $\mathcal{E}$ be defined by~\eqref{Def_threshold_energy_density}. Then, $\mathcal{E}\in L^1(\R) \cap L^\infty(\R)$ and
	\[
		\mathcal{E}(x) \geq (m+\omega) \psi_2^\infty(x)^2 + g(x) \psi_1^\infty(x)^2 > 0\,, \quad\forall\, x\in\R\,.
	\]
\end{lemma}
\begin{proof}
	Using the definition of~$W$, we rewrite
	\begin{equation}\label{Eq_formula_E}
		\mathcal{E}(x) = (2m - g(x)) \psi_2^\infty(x)^2 + g(x)\psi_1^\infty(x)^2\,.
	\end{equation}
	Recalling the definition of~$g$ in~\eqref{Formula_g}, we have
	\[
		0 < g(x) \leq g(0) = m-\omega\,, \quad\forall\, x\in\R\,,
	\]
	and consequently
	\[
		2m - g(x) \geq 2m - g(0) = m+\omega\,, \quad\forall\, x\in\R\,,
	\]
	which yields
	\[
		\mathcal{E}(x) \geq (m+\omega)\psi_2^\infty(x)^2 + g(x)\psi_1^\infty(x)^2 \geq g(x)|\psi^\infty (x)|^2>0\,, \quad\forall\, x\in\R\,,
	\]
	since $|\psi^\infty|^2 >0$ on~$\R$ by Proposition~\ref{bas1}.

	Finally, observing that $\psi_1^\infty \in L^\infty(\R)$, $\psi_2^\infty \in L^2(\R) \cap L^\infty(\R)$ by Proposition~\ref{Proposition_Property_resonances_at_thresholds} again, and $g \in L^1(\R) \cap L^\infty(\R)$ by~\eqref{Formula_g}, we conclude from~\eqref{Eq_formula_E} that~$\mathcal{E} \in L^1(\R) \cap L^\infty(\R)$.
\end{proof}

The following result will be used later when estimating commutators involving the resonance cutoff. It also relies on the fact that $\psi_2^\infty \in L^2(\R)$.
\begin{lemma}\label{Lemma_psi_hat}
	Let $(p,\omega)\in(0,+\infty)\times(0,m)$ and $\psi^\infty \equiv \psi_+^\infty \in L^\infty(\R,\Cbb^2)$ be a threshold resonance of~$A_q$ at $m$. Then, the function
	\[
		\Psi^\infty := (A_q - m)\, W \sigma_3 \psi^\infty,
	\]
	defined in the sense of distributions, belongs to $L^2(\R,\Cbb^2)$.
\end{lemma}

\begin{proof} We recall the formula
	\begin{equation*}
		A_q = i q \sigma_2 \partial_x + m\sigma_3 - (q+1) g \sigma_3
	\end{equation*}
	and compute
	\begin{align*}
		(A_q-m) W \sigma_3 \psi^\infty &= iq\sigma_2\sigma_3\, \partial_x (W \psi^\infty) + W \sigma_3 \big(m(\sigma_3 - \Id) - (q+1)g\sigma_3\big)\psi^\infty \\
			&=\begin{multlined}[t][0.7\textwidth]
				-qW' \sigma_1 \psi^\infty - W \sigma_3 (iq\sigma_2\partial_x)\psi^\infty\\
				+ W \sigma_3 \big(m(\sigma_3 - \Id) - (q+1)g\sigma_3\big)\psi^\infty.
			\end{multlined}
	\end{align*}
	Using the resonance equation
	\[
		iq\sigma_2\partial_x\psi^\infty = -\big(m( \sigma_3-\Id) -(q+1)g\sigma_3\big)\psi^\infty,
	\]
	we obtain
	\begin{align*}
		(A_q-m) W \sigma_3 \psi^\infty &= -q\sigma_1 W' \psi^\infty + 2 W \sigma_3 \big(m(\sigma_3 - \Id) - (q+1)g\sigma_3\big)\psi^\infty \\
			&= q g' \sigma_1 \psi^\infty + 2m W (\Id - \sigma_3)\psi^\infty - 2(q+1) g W \psi^\infty.
	\end{align*}
	In particular,
	\reversemarginpar%
	\begin{multline*}
		\norm{ (A_q-m) W \sigma_3 \psi^\infty }_2 \leq q\norm{g'}_2 \norm{\psi^\infty}_\infty + 2m \norm{W}_\infty \norm{ (\Id - \sigma_3)\psi^\infty }_2\\
		+ 2(q+1) \norm{W}_\infty \norm{\psi^\infty}_\infty \norm{g}_2\,,
	\end{multline*}
	where $g$ and $g' = -2\omega u_1 v_1$ belong to $L^2(\R)$, since they are continuous functions that decay exponentially (see, for instance,~\cite[Appendix A]{AldRicStoVDB-23}). Moreover, $\norm{W}_\infty = m$ and the resonance~$\psi^\infty \in L^\infty(\R)$, so the first and third terms are bounded. For the second term, we use Proposition~\ref{Proposition_Property_resonances_at_thresholds} to see that
	\[
		(\Id - \sigma_3)\psi^\infty = (0, 2\psi_2^\infty)\transp \in L^2(\R,\Cbb^2)\,.
	\]
	 This proves that $\Psi^\infty \in L^2(\R,\Cbb^2)$.
\end{proof}

\subsection{Properties of the trial states.}
In order to apply the min-max principle, we need admissible trial functions in~$H^1(\R,\Cbb^2)$. Since the threshold resonance $\psi^\infty$ does not belong to $L^2(\R,\Cbb^2)$, we construct such trial states by localizing $\psi^\infty$ with an exponential cutoff. 
For any $\delta>0$, we define the trial state
\begin{equation}\label{trialstate}
	\psi_\delta = \sqrt{\delta}\, e^{-\delta|\cdot|} \psi^\infty =: \eta_\delta \psi^\infty\,.
\end{equation}
Then, $\norm{\eta_\delta}_2 = 1$. Moreover, $\psi_\delta \in H^1(\R,\Cbb^2)$ and, using~\eqref{def_c_infty},
\[
	c_\infty \leq \norm{\psi_\delta}_2^2 \leq \norm{\psi^\infty}_\infty^2\,.
\]

%\begin{remark}\label{rem:1}
%	Recall that the eigenfunction $\phi_0 = (v,u)\transp$ spans the eigenspace $E_{\{\omega\}}(A_q)$, and $\sigma_p(A_q)\cap(-\omega,\omega)=\emptyset$, see~\cite[Theorem 1.3]{AldRicStoVDB-23}. Therefore, defining
%	\[
%		\Lambda_-^{\circ} := E_{(-\infty,-\omega]}(A_1) = E_{(-\infty,+\omega)}(A_1)\,,
%	\]
%	we have
%	\begin{equation}\label{sim}
%		\Lambda_- \psi_\delta = \Lambda_-^\circ \psi_\delta + E_{\{\omega\}}(A_1)\psi_\delta = \Lambda_-^\circ \psi_\delta\,.
%	\end{equation}
%	Indeed, $\pscal{\phi_0}{\psi_\delta} = 0$ by parity: $v$ is even while the first component of~$\psi_\delta$ is odd, and $u$ is odd while the second component of~$\psi_\delta$.
%\end{remark}
The following fundamental properties of the trial states will be used repeatedly.

\begin{lemma}\label{lem:trial}
	Let $\delta>0$, $q > 0$, and $\psi_\delta=\eta_\delta \psi^\infty\in H^1(\R,\Cbb^2)$ be defined as above.
	\begin{enumerate}
		\item The function $\psi_\delta$ satisfies
		\[
			A_q\psi_\delta = (m - iq\delta \sigma_2 \sgn(\cdot))\psi_\delta \quad \text{pointwise on } \R
		\]
		and, in particular,
		\begin{equation}\label{energy}
			\pscal{\psi_\delta}{A_q \psi_\delta } = m \norm{\psi_\delta}_2^2\,.
		\end{equation}
		
		\item The projection onto $F_-$ satisfies the estimate
		\[
			\norm{ \Lambda_- \psi_\delta }_2 \leq \frac{q\delta}{m-\tau} \norm{\psi_\delta}_2\,.
		\]
	\end{enumerate}
\end{lemma}

\begin{proof}
	We first compute, in the sense of distributions
	\[
		A_q \psi_\delta = m\psi_\delta + iq \sigma_2 \eta'_\delta \psi^\infty = (m - iq \delta \sigma_2 \sgn(\cdot))\psi_\delta\,,
	\]
	where we used that $\psi^\infty$ is a resonance of~$A_q$ at $m$.
	
	Since $A_q$ has real coefficients we can always assume $\psi^\infty$ real valued and consequently $\psi_\delta = (\psi_{\delta,1}, \psi_{\delta,2})\transp \in \R\times \R $. Thus, 
	\[
		\pscal{\psi_\delta}{i \sigma_2 \sgn(\cdot)\psi_\delta} = \pscal{\psi_{\delta,1}}{\sgn(\cdot)\psi_{\delta,2}} + \pscal{\psi_{\delta,2}}{-\sgn(\cdot)\psi_{\delta,1}} = 0\,,
	\]
	and therefore 
	\[
		\pscal{\psi_\delta}{A_q\psi_\delta} = m \norm{\psi_\delta}_2^2 \,,
	\]
	which proves~\eqref{energy}.
	
	For the second point, by the definition of~$\Lambda_-$ in~\eqref{Def_Lambda_pm_NearBelow_1}, we write
	\begin{align*}
		\norm{ \Lambda_- \psi_\delta }_2 = \norm{ (A_q-m)^{-1}\Lambda_- (A_q-m)\psi_\delta }_2 &=q \delta \norm{ (A_q-m)^{-1}\Lambda_- \sgn(\cdot)\sigma_2 \psi_\delta }_2 \\
			&\leq q\delta \norm{ (A_q-m)^{-1}\Lambda_- } \norm{\psi_\delta}_2 \leq \frac{q\delta}{m-\tau} \norm{\psi_\delta}_2\,,
	\end{align*}
	where the last inequality follows from the spectral properties encoded in~$\Lambda_-$.
\end{proof}
\subsection{Strategy.}
We apply the min--max principle to $A_p$ with the projections $\Lambda_\pm$ defined in~\eqref{Def_Lambda_pm_NearBelow_1}. Recall that, for $\varepsilon>0$ sufficiently small, the gap condition holds and
\[
\gamma_0 < (1-2\varepsilon)m \leq \gamma_1\,.
\]
Our goal is therefore to prove that
\[
\gamma_1 < m = \gamma_\infty\,.
\]
The min--max principle then guarantees that $\gamma_1$ is an eigenvalue of~$A_p$ in~$(\gamma_0,m)$.
\begin{remark}
	Let us verify that $\gamma_1$ corresponds to a new eigenvalue in the gap, rather than to the continuation of the eigenvalue $\tau$ of~$A_q$. Let $\lambda_\tau(p)=\lambda_\tau((1-\varepsilon) q)$ denote the eigenvalue branch satisfying $\lambda_\tau(q)=\tau$. By analyticity of the eigenvalue branches in the gap (see the introduction of Section~\ref{Section_proof_prop_p_less_than_1}),
	\[
	\lambda_\tau(p)=\tau+o(1)\,.
	\]
	Since $\tau\in[\omega,m)$, for $\varepsilon>0$ sufficiently small,
	\[
	\lambda_\tau(p)<(1-2\varepsilon)m\leq\gamma_1\,.
	\]
	If $\lambda_\tau(p)\leq\gamma_0$, it clearly cannot coincide with $\gamma_1$. If instead $\lambda_\tau(p)>\gamma_0$, the min--max principle shows that $\lambda_\tau(p)$ is an eigenvalue preceding $\gamma_1$. In either case, $\gamma_1$ is distinct from the continuation of $\tau$ and therefore corresponds to a new eigenvalue in the gap.
\end{remark}

\medskip

In order to prove   $\gamma_1 < m$, we use the trial states $\psi_\delta$ constructed above. We have
\[
	\norm{\Lambda_+ \psi_\delta}_2^2 = \norm{\psi_\delta}_2^2 - \norm{\Lambda_- \psi_\delta}_2^2 \geq \left(1 - \frac{q^2 \delta^2}{(m-\tau)^2} \right) \norm{\psi_\delta}_2^2
\]
by Lemma~\ref{lem:trial}, so that $\Lambda_+ \psi_\delta \neq 0$ for~$\delta < (m-\tau)/q$. Hence, using the test function $\varphi_+ = \Lambda_+ \psi_\delta \neq 0$ in the min-max characterization~\eqref{gap-con} of~$\gamma_1$, we obtain
\[
	\gamma_1 = \inf_{\varphi_+ \in F_+\setminus\{0\}} \sup_{\varphi \in (\vect\{\varphi_+\} \oplus F_-) \setminus \{0\}} \frac{\langle \varphi, A_p \varphi \rangle}{ \norm{\varphi}_2^2} \leq \sup_{\varphi \in (\vect\{\Lambda_+ \psi_\delta\} \oplus F_-) \setminus \{0\}} \frac{\langle \varphi, A_p \varphi \rangle}{ \norm{\varphi}_2^2}\,.
\]
%and since the supremum is taken over all $\varphi_- \in F_-$, we can replace $\Lambda_+ \psi_\delta$ by $\psi_\delta$:
Using that $\vect\{\Lambda_+ \psi_\delta\} \oplus F_- = \vect\{\psi_\delta\} + F_-$, we compute the r.h.s.~as
\begin{align*}
	\text{r.h.s.} &= \sup_{\varphi \in (\vect\{\psi_\delta\} + F_-) \setminus \{0\}} \frac{\langle \varphi, A_p \varphi \rangle}{ \norm{\varphi}_2^2} = \sup_{\substack{\alpha \in \Cbb, \ \varphi_- \in F_-\\(\alpha, \varphi_-) \neq (0, 0)}} \frac{ \pscal{\alpha \psi_\delta + \varphi_-}{A_p(\alpha \psi_\delta + \varphi_-)} }{ \norm{ \alpha \psi_\delta + \varphi_- }_2^2 } \\
		&= \max \left\{ \sup_{\varphi_- \in F_-\setminus\{0\}} \frac{ \pscal{\varphi_-}{A_p\varphi_-} }{ \norm{ \varphi_- }_2^2 }, \ \sup_{\substack{\alpha \in \Cbb\setminus\{0\}\\\varphi_- \in F_-}} \frac{ \pscal{\psi_\delta + \alpha^{-1}\varphi_-}{A_p(\psi_\delta + \alpha^{-1}\varphi_-)} }{ \norm{ \psi_\delta + \alpha^{-1}\varphi_- }_2^2 } \right\} \\
		&= \max \left\{ \gamma_0, \ \sup_{\varphi_- \in F_-} \frac{ \pscal{\psi_\delta + \varphi_-}{A_p(\psi_\delta + \varphi_-)} }{ \norm{ \psi_\delta + \varphi_- }_2^2 } \right\} = \sup_{\varphi_- \in F_-} \frac{ \pscal{\psi_\delta + \varphi_-}{A_p(\psi_\delta + \varphi_-)} }{ \norm{ \psi_\delta + \varphi_- }_2^2 }\,,
\end{align*}
where the last equality is a consequence of our smallness assumption on $\varepsilon$, which ensures that $\gamma_0 < \gamma_1$ (see Subsection~\ref{Subsect_Admissible_proj}), hence preventing $\text{r.h.s.} = \gamma_0$ as it would yield the contradiction $\gamma_1 \leq \gamma_0$.
We are thus reduced to proving that
\[
	\sup_{\varphi_- \in F_-} \boldsymbol{\gamma}(\varphi_-) < m\,,
\]
where we introduce the functional
\begin{equation}\label{gama-bold}
	\boldsymbol{\gamma}(\varphi_-) := \frac{ \pscal{\psi_\delta}{A_p\psi_\delta} + 2\Re \pscal{\varphi_-}{A_p\psi_\delta} + \pscal{\varphi_-}{A_p\varphi_-} }{ \norm{ \psi_\delta + \varphi_- }_2^2 }\,.
\end{equation}

\medskip
The proof proceeds by estimating separately the four contributions appearing in~\eqref{gama-bold}:
\begin{itemize}
	\item[--] the \emph{main term} $\pscal{\psi_\delta}{A_p \psi_\delta}$, which yields a negative contribution of order $\varepsilon\delta$; see Lemma~\ref{lem:first-term}\,;
	
	\item[--] the denominator, controlled in Lemma~\ref{lem-norm}\,;
	
	\item[--] the \emph{cross terms}, estimated in Corollary~\ref{cor:cross}\,;
	
	\item[--] the \emph{purely $F_-$ term}, $\pscal{\varphi_-}{A_p\varphi_-}$, controlled through the bound~\eqref{UpperBound_gamma0} on~$\gamma_0$\,.
\end{itemize}
We then conclude by choosing $\delta$ as a suitable function of~$\varepsilon$.

\subsection{Key estimates.}
We begin by
 estimating the main contribution in the numerator, which provides the key energy drop.
\begin{lemma}\label{lem:first-term}
	Let $\delta>0$. Then, 
	\[
		\pscal{\psi_\delta}{A_p\psi_\delta} \leq (m-\varepsilon \delta E_\delta) \norm{\psi_\delta}_2^2\,,
	\]
	where 
	\[
		E_\delta := \frac{1}{ \norm{\psi^\infty}_\infty^2}\int_\R e^{-2\delta |x|} \mathcal{E}(x) \di x \underset{\delta\to0}{\longrightarrow} \frac{ \norm{\mathcal{E}}_1}{ \norm{\psi^\infty}_\infty^2} =: E_\star > 0\,.
	\]
\end{lemma}
\begin{proof}
	Using the identities in~\eqref{energy}, \eqref{obs}, and \eqref{Def_threshold_energy_density}, we get 
	\begin{align*}
		{\pscal{\psi_\delta}{A_p \psi_\delta}} &= m \norm{\psi_\delta}_2^2 - \varepsilon\Big(m \norm{\psi_\delta}_2^2 - \pscal{\psi_\delta}{W\sigma_3 \psi_\delta}\Big) = m \norm{\psi_\delta}_2^2 - \varepsilon \int_\R \mathcal{E}(x) \eta^2_\delta(x)\di x \\
		&= m \norm{\psi_\delta}_2^2 - \varepsilon \delta \int_\R e^{-2\delta |x|} \mathcal{E}(x) \di x = m \norm{\psi_\delta}_2^2 - \varepsilon \delta E_\delta \norm{\psi^\infty}_\infty^2\,,
	\end{align*}
	which gives the upper bound since $\norm{\psi_\delta}_2 \leq \norm{\psi^\infty}_\infty$. Moreover, thanks to Lemma~\ref{lem:energy-density}, we have that $\mathcal{E}\in L^1$ holds. Therefore, by the theorem of dominated convergence, it follows that $E_\delta$ converges to the claimed value as~$\delta\to 0$. Finally, the positivity of~$\mathcal{E}$ established in Lemma~\ref{lem:energy-density} ensures that $E_\star > 0$.
\end{proof}
\begin{remark*}
	Note that in the case~$p=1$ one can compute $E_\star$ explicitly and therefore verify its positivity. Indeed, since% $\norm{\mathcal{E}}_1$ and $\norm{\psi^\infty}_\infty$ are
	\[
		\norm{\psi^\infty}_\infty = \frac{\sqrt{m^2-\omega^2}}{2\omega} \quad \text{and} \quad \norm{\mathcal{E}}_1 = \frac{m^2-\omega^2}{(2\omega)^2} \ln\left( \frac{m+\sqrt{m^2-\omega^2}}{\omega} \right)\,,
	\]
	thence
	\[
		E_\star = \frac{\sqrt{m^2-\omega^2}}{2\omega} \ln\left( \frac{m+\sqrt{m^2-\omega^2}}{\omega} \right) >0 \,.
	\]
\end{remark*}
Next we control the denominator in~\eqref{gama-bold}.
\begin{lemma}\label{lem-norm}
	Let $\delta>0$. Then,
	\[
		\norm{ \varphi + \psi_\delta }_2^2 \leq \left( 1+\frac{q\delta}{m-\tau} \right) \left( \norm{\varphi}_2^2 + \norm{\psi_\delta}_2^2 \right)\,, \quad \forall\, \varphi \in F_-\,.
	\]
\end{lemma}	
\begin{proof}
	We get the desired result using Lemma~\ref{lem:trial}, since
	\begin{align*}
		\norm{ \varphi + \psi_\delta }_2^2 &\leq \norm{\varphi}_2^2 + \norm{\psi_\delta}_2^2 + 2\left|\pscal{\varphi}{\psi_\delta}\right| = \norm{\varphi}_2^2 + \norm{\psi_\delta}_2^2 + 2\left|\pscal{\Lambda_- \varphi}{\psi_\delta}\right| \\
		&\leq \norm{\varphi}_2^2 + \norm{\psi_\delta}_2^2 + 2 \norm{\varphi}_2 \norm{ \Lambda_- \psi_\delta }_2 \\
		&\leq \norm{\varphi}_2^2 + \norm{\psi_\delta}_2^2 + 2\frac{q\delta}{m-\tau} \norm{\varphi}_2 \norm{\psi_\delta}_2 \\
		&\leq \left( 1+\frac{q\delta}{m-\tau} \right) \left( \norm{\varphi}_2^2 + \norm{\psi_\delta}_2^2 \right) \,. \qedhere
	\end{align*}
\end{proof}
It remains to control the cross terms appearing in~\eqref{gama-bold}. The following bound is an immediate consequence of Lemmas~\ref{lem:1-cross-term} and~\ref{lem:2-cross-term} proved below.
\begin{corollary}\label{cor:cross}
	Let $q > 0$, $C:=q\frac{2m-\tau}{2(m-\tau)}>0$, $\delta \in (0, (qm)^{-2}]$, $\varepsilon \in (0,1)$, and $\alpha\in \R$. Then,
	\begin{multline*}
		2 {\Re}\,\pscal{\varphi}{A_p \psi_\delta} \leq 2 m\Re\pscal{\varphi}{\psi_\delta} + \left( \varepsilon^2 \delta^{1-\alpha} \left( 1 + c_\infty^{-1} \normSM{\Psi^\infty}_2^2 \right) + 2 C \delta^{2-\alpha} \right) \norm{\psi_\delta}_2^2 \\
		+ 2 \delta^{\alpha} \left(C+\frac{1}{(m-\tau)^2}\right) \norm{\varphi}_2^2\,, \qquad \forall\, \varphi \in F_-\,.
	\end{multline*}
\end{corollary}
\begin{lemma}\label{lem:1-cross-term} Let $q > 0$, $\delta \in (0, (qm)^{-2}]$, and $\alpha\in\R$. Then,
	\[
		2 \varepsilon \left|\pscal{\varphi}{W \sigma_3 \psi_\delta}\right| \leq \varepsilon^2 \delta^{1-\alpha} \left( 1 + c_\infty^{-1} \normSM{\Psi^\infty}_2^2 \right) \norm{\psi_\delta}_2^2 + \frac{2 \delta^\alpha}{(m-\tau)^2} \norm{\varphi}_2^2\,, \quad \forall\, \varphi \in F_-\,.
	\]
	where $c_\infty>0$ is defined in~\eqref{def_c_infty} and $\Psi^\infty$ is the $L^2$-function defined in Lemma~\ref{Lemma_psi_hat}.
\end{lemma}
\begin{proof}
	For all $\varphi \in F_-$ we have 
	\begin{align*}
		\left|\pscal{\varphi}{\sigma_3 W\psi_\delta}\right| = \left|\pscal{\Lambda_-\varphi}{\sigma_3 W\psi_\delta}\right| &\leq \norm{ (A_q -m)^{-1} \Lambda_-\varphi }_2 \norm{(A_q -m)\sigma_3 W\psi_\delta}_2 \\
		&\leq \frac{ \norm{\varphi}_2 }{m-\tau} \norm{(A_q -m)\sigma_3 W\psi_\delta}_2\,,
	\end{align*}
	since $A_q -m$ is self-adjoint.
	Next, we recall Lemma~\ref{Lemma_psi_hat} and calculate in the sense of distributions,
	\begin{align*}
		(A_q -m)\sigma_3 W\psi_\delta = (A_q -m)\eta_\delta \sigma_3 W\psi^\infty &= \eta_\delta (A_q -m) \sigma_3 W\psi^\infty+[A_q,\eta_\delta] \sigma_3 W\psi^\infty\\
			&=\eta_\delta \Psi^\infty-iq\delta \sigma_2\sgn(\cdot) W \sigma_3 \psi_\delta\,.
	\end{align*}
	Thus, 
	\begin{align*}
		\norm{(A_q -m)\sigma_3 W\psi_\delta}_2 \leq q \delta \norm{W} \norm{\psi_\delta}_2 + \norm{\eta_\delta}_\infty \normSM{\Psi^\infty}_2 &= q\delta m \norm{\psi_\delta}_2 + \delta^{1/2} \normSM{\Psi^\infty}_2 \\
			&\leq \delta^{1/2}( \norm{\psi_\delta}_2 + \normSM{\Psi^\infty}_2)\,,
	\end{align*}
	since $qm \leq \delta^{-1/2}$ by assumption. Therefore, for any $\alpha\in\R$, we get
	\begin{align*}
		2\varepsilon \left|\pscal{\varphi}{\sigma_3 W\psi_\delta}\right|&\leq \,2\varepsilon\delta^{\frac{1-\alpha}{2}} \norm{\psi_\delta}_2 \,\frac{\delta^{\alpha/2} \norm{\varphi}_2 }{m-\tau} + 2\varepsilon \delta^{\frac{1-\alpha}{2}} \normSM{\Psi^\infty}_2 \frac{\delta^{\alpha/2} \norm{\varphi}_2 }{m-\tau}\\
		&\leq \varepsilon^2\delta^{1-\alpha} \left( \norm{\psi_\delta}_2^2 + \normSM{\Psi^\infty}_2^2 \right)+\frac{2 \delta^\alpha}{(m-\tau)^2} \norm{\varphi}_2^2\,,
	\end{align*}
	for all $\varphi \in F_-$. We conclude the proof recalling that $\norm{\psi_\delta}_2^2 \geq c_\infty$.
\end{proof} 

The remaining cross term in Corollary~\ref{cor:cross} is considered in the following.
\begin{lemma}\label{lem:2-cross-term}
	Let $q > 0$, $C:=q\frac{2m-\tau}{2(m-\tau)}>0$, and $\varepsilon \in (0,1)$. Then,
	\[
		(1-\varepsilon)\Re\pscal{\varphi}{A_q \psi_\delta} \leq m\Re\pscal{\varphi}{\psi_\delta} + C\Par{{\delta^{\alpha}} \norm{\varphi}_2^2 + \delta^{2-\alpha} \norm{\psi_\delta}_2^2}\,, \quad \forall\, \varphi \in F_-\,.
	\]
\end{lemma}
\begin{proof}
	We consider $R := (1-\varepsilon)\Re\pscal{\varphi}{A_q \psi_\delta} - m \Re\pscal{\varphi}{\psi_\delta}$. By Lemma~\ref{lem:trial}, we have
	\begin{align*}
		R &= -\varepsilon m \Re\pscal{\varphi}{\psi_\delta} + (1-\varepsilon) \Re\pscal{\varphi}{-iq\delta\sigma_2 \sgn(\cdot) \psi_\delta} \\
		&\leq m \left| \pscal{\varphi}{\Lambda_- \psi_\delta} \right| + q\delta \left| \pscal{\varphi}{-i\sigma_2 \sgn(\cdot) \psi_\delta} \right| \\
		&\leq \norm{\varphi}_2 \left(m \norm{\Lambda_- \psi_\delta}_2 + q\delta \norm{\psi_\delta}_2 \right) \\
		&\leq q\delta \norm{\varphi}_2 \left( \frac{m}{m-\tau} + 1 \right) \norm{\psi_\delta}_2 = 2 C \delta \norm{\varphi}_2 \norm{\psi_\delta}_2 = 2 C \delta^{\frac{\alpha}{2}} \norm{\varphi}_2 \delta^{1 - \frac{\alpha}{2}} \norm{\psi_\delta}_2 \\
		&\leq C\left(\delta^{\alpha} \norm{\varphi}_2^2 + \delta^{2 - \alpha} \norm{\psi_\delta}_2^2\right) \,,
	\end{align*}
	for all $\varphi \in F_-$ and $\varepsilon \in (0,1)$.
\end{proof}

\medskip
	
\subsection{Conclusion of the proof}
We now combine the previous estimates and choose the parameters appropriately to conclude the proof. 
The key point is that the negative contribution of order $\varepsilon\delta$ dominates all remaining terms.

\medskip

Let $\alpha \in (0,1/2)$ and $\varepsilon>0$ be sufficiently small, and set
\[
	\delta = \varepsilon^{1+2\alpha}\,.
\]
By Lemma~\ref{lem:first-term}, the main term produces a negative contribution of order
\[
	\varepsilon\delta = \varepsilon^{2(\alpha+1)}\,.
\]
For $\alpha \in (0,1/2)$, the remaining terms estimated in Corollary~\ref{cor:cross} satisfy
\[
	\varepsilon^2 \delta^{1-\alpha} = \varepsilon \delta\, o(1), \qquad \delta^{2-\alpha} = \varepsilon \delta\, o(1), \qquad \text{and} \qquad \delta^\alpha = o(1)\,,
\]
as $\varepsilon \to 0$\,.
	
\medskip
	
Combining~\eqref{UpperBound_gamma0}, Lemma~\ref{lem:first-term}, Corollary~\ref{cor:cross}, and~\eqref{gama-bold}, we obtain, for all $\varphi \in F_-$, that
\[
	\norm{ \varphi + \psi_\delta }_2^2 \boldsymbol{\gamma}(\varphi) \leq \big(m - \varepsilon \delta E_\star + \varepsilon \delta o(1)\big) \norm{\psi_\delta}_2^2 + 2 m\Re\pscal{\varphi}{\psi_\delta} + \big( \tau + o(1)\big) \norm{\varphi}_2^2\,,
\]
as $\varepsilon \to 0$\,.

Recalling that $\tau\in [\omega, m)$, we choose $\varepsilon>0$ sufficiently small so that
\[
	\tau + o(1) < m - \varepsilon\delta E_\star + o(\varepsilon\delta) < m\,,
\]
and therefore infer that
\begin{multline*}
	\norm{ \varphi + \psi_\delta }_2^2 \boldsymbol{\gamma}(\varphi) \leq \big(m - \varepsilon\delta (E_\star + o(1))\big) \big( \norm{\psi_\delta}_2^2 + \norm{\varphi}_2^2 \big) + 2 m\Re\pscal{\varphi}{\psi_\delta} \\
	= m \norm{ \varphi + \psi_\delta }_2^2 - \varepsilon\delta (E_\star + o(1)) \big( \norm{\psi_\delta}_2^2 + \norm{\varphi}_2^2 \big)\,,
\end{multline*}
for all $\varphi \in F_-$.

\medskip

Using Lemma~\ref{lem-norm}, we deduce for all $\varphi \in F_-$ that
\[
	\boldsymbol{\gamma}(\varphi) \leq m - \varepsilon\delta (E_\star + o(1)) \frac{\norm{\psi_\delta}_2^2 + \norm{\varphi}_2^2}{\norm{\varphi + \psi_\delta}_2^2} \leq m - \varepsilon\delta (E_\star + o(1)) \left(1 + \frac{q \delta}{m-\tau}\right)^{-1} < m\,,
\]
for~$\varepsilon$ sufficiently small and with $\delta = \varepsilon^{1+2\alpha}$ for any $\alpha\in(0, 1/2)$.

\medskip

Therefore,
\[
	\sup_{\varphi \in F_-} \boldsymbol{\gamma}(\varphi) < m\,,
\]
and the min-max principle implies that $A_p$ admits $\gamma_1$ as eigenvalue in~$(\omega,m)$. Moreover, we obtained the bounds
\[
	m - 2m\varepsilon \leq \gamma_1 \leq m - E_\star \varepsilon^{2+s} + o\!\left(\varepsilon^{2+s} \right)
\]
for all $s:=2\alpha \in(0, 1)$ as claimed.
By symmetry of the spectrum, the same holds in~$(-m,-\omega)$, which concludes the proof of Proposition~\ref{prop-birth}.

\section{Simplicity of threshold generalized eigenvalues}\label{Section_proof_main_thm_3_Resonances}
We prove Theorem~\ref{Thm_at_most_one_generalized_eigenfunction_at_thresholds}, and hence Theorem~\ref{main_thm_3_Resonances}, by combining two ingredients. First, Proposition~\ref{Proposition_at_most_one_generalized_eigenfunction_at_thresholds_conditional} below establishes a conditional simplicity criterion for generalized eigenvalues of one-dimensional Dirac operators with $L^1_{\rm loc}(\R)$ potentials. Roughly speaking, it shows that a generalized eigenvalue is simple whenever all associated generalized eigenfunctions share a common component vanishing at the same point.

Second, Proposition~\ref{Proposition_Property_resonances_at_thresholds}, proved in Section~\ref{Section_proof_main_thm_2_noEV}, shows that generalized eigenfunctions at the thresholds $\lambda=\pm m$ possess precisely this property when
\[
V \in L^1\cap L^2(\R,\Cbb^{2\times2})\,.
\]
Indeed, at each threshold one component necessarily belongs to $H^1(\R)$ and therefore vanishes at infinity. The conditional simplicity criterion can then be applied, yielding the simplicity of generalized eigenvalues at the thresholds.

The remainder of this section is devoted to the proof of the general conditional simplicity result presented above and to some of its consequences. We begin by introducing a structural assumption on the potential under which the argument applies.
	
\begin{assumption}\label{Potential_sigma_2_condition}
	Let $V: \R \to \Cbb^{2\times2}$ be such that its decomposition
	\[
		V(x)=\alpha_0 (x) \Id+\alpha_1 (x)\sigma_1 +\alpha_2 (x) \sigma_2 + \alpha_3 (x) \sigma_3
	\]
	is such that $\alpha_2$ is real-valued.
\end{assumption}
	
\begin{remark*}\leavevmode
	This assumption is equivalent to the assumption in Theorem~\ref{Thm_at_most_one_generalized_eigenfunction_at_thresholds} that $V-V\transp$ is purely imaginary-valued. In particular, any symmetric potential ($V=V\transp$) satisfies Assumption~\ref{Potential_sigma_2_condition}.
\end{remark*}	
%\begin{remark*}\leavevmode
%	\begin{itemize}
%		\item This assumption is equivalent to the assumption in Theorem~\ref{Thm_at_most_one_generalized_eigenfunction_at_thresholds} that $V-V\transp$ is purely imaginary-valued.
%		
%		\item In particular, any symmetric potential ($V=V\transp$) satisfies Assumption~\ref{Potential_sigma_2_condition}.
%		
%		\item In Theorem~\ref{Thm_at_most_one_generalized_eigenfunction_at_thresholds}, the potential is already assumed to belong to $L^1(\R)$. Therefore, when applying the present assumption in that theorem, the only additional requirement is that $\alpha_2$ be real-valued.
%		
%		\item We formulate Assumption~\ref{Potential_sigma_2_condition} with the weaker condition $\alpha_2\in L^1_{\rm loc}(\R)$ in order to keep Proposition~\ref{Proposition_at_most_one_generalized_eigenfunction_at_thresholds_conditional} and Corollary~\ref{Corollary_simplicity_generalized_eigenfunction_under_parity} applicable in a more general setting.
%	\end{itemize}
%\end{remark*}

\begin{proposition}[Conditional simplicity of generalized eigenvalues]\label{Proposition_at_most_one_generalized_eigenfunction_at_thresholds_conditional}\leavevmode\nopagebreak

	Let $V \in L^1_{\rm loc}(\R,\Cbb^{2\times2})$ satisfy Assumption~\ref{Potential_sigma_2_condition} and let $\lambda\in\Cbb$. If there exist $i\in\{1,2\}$ and $x_0\in\R\cup\{\pm\infty\}$ such that any generalized eigenfunction $\Psi=(\psi_1,\psi_2)$ of~$D_m + V$ associated to $\lambda$ satisfies $\psi_i(x_0) = 0$, then
	\[
		\dim \left\{ \text{generalized eigenfunction of~$D_m + V$ associated to $\lambda$} \right\} \leq 1\,.
	\]
\end{proposition}
\begin{remark*}\leavevmode
	\begin{itemize}
		\item When $x_0\in\{\pm\infty\}$, ``$\psi_i(\pm\infty)$'' means $\lim_{\pm\infty}\psi_i$. As before, a generalized eigenfunction $\Psi$ of~$D_m + V$ associated with $\lambda$ is a distributional solution $\Psi \in L^\infty(\R,\Cbb^2)$ of~$(D_m + V)\Psi = \lambda \Psi$.
		
		\item Assumption~\ref{Potential_sigma_2_condition} is crucial to our argument. 
			The fact that $\alpha_2$ is real-valued allows one to perform a unitary gauge transformation that eliminates the $\sigma_2$ component of the potential, reducing the problem to the case of a symmetric potential $V=V\transp$.
		
		\item It is also crucial that the index $i$ and the point $x_0$ are the same for \emph{all} generalized eigenfunctions associated with the given generalized eigenvalue. 
			This condition ensures that the Wronskian used in the proof vanishes.
		
		\item Finally, we stress that this proposition applies to any (complex) generalized eigenvalue. 
			In particular, it is not restricted to real eigenvalues and does not rely on the threshold structure.
	\end{itemize}
\end{remark*}

Before turning to the proof of Proposition~\ref{Proposition_at_most_one_generalized_eigenfunction_at_thresholds_conditional}, we describe an immediate consequence concerning the number of generalized eigenfunctions belonging to the parity subspaces~$L^\infty_\pm$ defined in \ref{Def_even_odd_function_subspaces}.

\begin{corollary}[Simplicity of \texorpdfstring{$L^\infty_\pm$}{L∞±}-generalized eigenvalues]\label{Corollary_simplicity_generalized_eigenfunction_under_parity}\leavevmode\nopagebreak

	Let $\lambda\in\Cbb$ and let $V \in L^1_{\rm loc}(\R, \Cbb^{2\times2})$ be a parity-preserving potential satisfying Assumption~\ref{Potential_sigma_2_condition}. Then,
	\[
		\dim\big\{\text{generalized eigenfunctions in~$L^\infty_\tau$ of~$D_m+V$ associated with $\lambda$}\big\} \leq 1\,,
	\]
	 for~$\tau\in\{+,-\}$.
	
	In particular, this holds for~$D_m+V=L_\mu$, $\mu\in\R$.
\end{corollary}
\begin{proof}
%	By definition of~$L^\infty_\pm$, any generalized eigenfunction in~$L^\infty_+$ (resp.~$L^\infty_-$) satisfies the condition in Proposition~\ref{Proposition_at_most_one_generalized_eigenfunction_at_thresholds_conditional} with $i=2$ (resp.~$i=1$) and $x_0=0$. Moreover, since both $D_m$ and $V$ preserve the parity decomposition $L^\infty(\R,\Cbb^2)=L^\infty_+\oplus L^\infty_-$, the operator $D_m+V$ leaves $L^\infty_\pm$ invariant. Therefore, Proposition~\ref{Proposition_at_most_one_generalized_eigenfunction_at_thresholds_conditional} applies to the restrictions $(D_m+V)_{\left|L^\infty_+\right.}$ and $(D_m+V)_{\left|L^\infty_-\right.}$, which proves the first claim.
	Due to the parity of functions in~$L^\infty_\pm$, any generalized eigenfunction in~$L^\infty_+$ (respectively $L^\infty_-$) satisfies the condition in Proposition~\ref{Proposition_at_most_one_generalized_eigenfunction_at_thresholds_conditional} with $i=2$ (resp.~$i=1$) and~$x_0=0$. Moreover, $D_m+V$ is parity-preserving since both $D_m$ and $V$ are, and applying Proposition~\ref{Proposition_at_most_one_generalized_eigenfunction_at_thresholds_conditional} respectively to the restrictions $(D_m+V)_{\left|\even(\R, \Cbb^2)\right.}$ and $(D_m+V)_{\left|\odd(\R, \Cbb^2)\right.}$ proves the first claim.
	
	The statement for~$L_\mu$ follows immediately, since the potential $V = L_\mu - D_m$ satisfies the assumptions.
\end{proof}
\begin{remark*}\leavevmode
	\begin{itemize}
		\item This result classifies the generalized eigenfunctions, of which there are at most two by standard ODE arguments: there is at most one in each of the parity subspaces.
		
		\item Note that this first consequence of Proposition~\ref{Proposition_at_most_one_generalized_eigenfunction_at_thresholds_conditional} is weaker than Theorem~\ref{Thm_at_most_one_generalized_eigenfunction_at_thresholds}. 
			Indeed, Theorem~\ref{Thm_at_most_one_generalized_eigenfunction_at_thresholds} asserts that there is at most one generalized eigenfunction at each threshold, whereas the present result only ensures that, for each threshold, there is at most one generalized eigenfunction in each of the subspaces~$L^\infty_+$ and~$L^\infty_-$. Nevertheless, this corollary provides a simple illustration of Proposition~\ref{Proposition_at_most_one_generalized_eigenfunction_at_thresholds_conditional}.
		
		\item Note that, in terms of the notation in Assumption~\ref{Potential_sigma_2_condition}, the parity-preserving property of~$V$ is equivalent to having $\alpha_0$ and $\alpha_3$ even functions from $\R$ to $\Cbb$, while $\alpha_1$ and $\alpha_2$ are odd functions.
	\end{itemize}
\end{remark*}
	
We now turn to the proof of Proposition~\ref{Proposition_at_most_one_generalized_eigenfunction_at_thresholds_conditional}. 
The argument is inspired by the proof of~\cite[Lemma~2.3]{AldRicStoVDB-23}, but it applies to a broader class of potentials and to generalized eigenfunctions instead of only eigenfunctions. 
%The key point is that the part of that proof establishing the constancy of the Wronskian only relies on the condition expressed in Assumption~\ref{Potential_sigma_2_condition}. 
%For completeness we reproduce the argument here and indicate where the additional assumption on the generalized eigenfunctions is used.
\begin{proof}[Proof of Proposition~\ref{Proposition_at_most_one_generalized_eigenfunction_at_thresholds_conditional}]
	Using the notation in Assumption~\ref{Potential_sigma_2_condition}, and since $\alpha_2$ is real-valued, $D_m+V$ is unitarily equivalent to
	\[
		U(x)\bigl(D_m+V(x)\bigr)U^*(x) = D_m+\alpha_0(x)\Id+\alpha_1(x)\sigma_1+\alpha_3(x)\sigma_3\,,
	\]
	where
	\[
		U(x)=\Ee{i\int_0^x \alpha_2(t)\,\di t}\,.
	\]
	Therefore, without loss of generality, we may restrict ourselves to symmetric potentials. That is, $V(x)=V(x)\transp$ or, equivalently, $V=\alpha_0\Id+\alpha_1\sigma_1+\alpha_3\sigma_3$.
	
	Since $V$ is symmetric, we can rewrite the (generalized) eigenvalue equation for~$\lambda$ as
	\begin{equation}\label{Proof_Proposition_at_most_one_generalized_eigenfunction_at_thresholds_conditional_Rewrite_Eigenvalu_Eq}
		\Psi'=\sum_{j=1}^3 \tilde\alpha_j\,\sigma_j\Psi\,,
	\end{equation}
	by defining the functions $\tilde\alpha_j:\R\to\Cbb$ as
	\[
		\tilde\alpha_1 := -(m+\alpha_3)\,, \qquad \tilde\alpha_2 := i (\alpha_0 - \lambda)\,, \qquad \text{and} \qquad \tilde\alpha_3 := \alpha_1\,.
	\]
	
	Assume now that $D_m+V$ admits two generalized eigenfunctions at $\lambda$. We denote them $\Phi=(\phi_1,\phi_2)$ and $\Xi=(\xi_1,\xi_2)$, and we first recall that they are absolutely continuous, by Proposition~\ref{bas1}. Let
	\[
		W(x):=\det\bigl(\Phi(x)\mid\Xi(x)\bigr)\,,
	\]
	which is also absolutely continuous. Hence, $W$ is differentiable a.e.~and we deduce from~\eqref{Proof_Proposition_at_most_one_generalized_eigenfunction_at_thresholds_conditional_Rewrite_Eigenvalu_Eq} that
	\begin{align*}
		W' &=\det(\Phi'\mid\Xi)+\det(\Phi\mid\Xi') \\
			&=\sum_{j=1}^3 \tilde\alpha_j \Bigl(\det(\sigma_j\Phi\mid\Xi)+\det(\Phi\mid\sigma_j\Xi)\Bigr)=0 \qquad \text{a.e.}\,,
	\end{align*}
	since
	\[
		\begin{aligned}[b]
			\det(\sigma_j\Phi\mid\Xi) =\det(\sigma_j\Phi\mid\sigma_j^2\Xi) &= \det(\sigma_j)\det(\Phi\mid\sigma_j\Xi)\\
				&=-\det(\Phi\mid\sigma_j\Xi)\,, &&\quad \forall\, j\in\{1,2,3\}\,.
		\end{aligned}
	\]
	Therefore, $W$ is constant on~$\R$ as an absolutely continuous function with a trivial derivative a.e.
	
	However, by assumption, either $\phi_1(x_0)=\xi_1(x_0)=0$ or $\phi_2(x_0)=\xi_2(x_0)=0$, which implies that $W(x_0)=0$. Hence, $W\equiv0$.
	Consequently, for each $x\in\R$, the vectors $\Phi(x)$ and $\Xi(x)$ are collinear in~$\Cbb^2$. 
	
	If $\Xi(0)\neq0$, then there exists $\alpha\in\Cbb$ such that $\Phi(0)=\alpha\Xi(0)$. 
	Since the equation is a linear first-order system with $L^1_{\rm loc}$-coefficients, Carathéodory's theorem guarantees the existence and uniqueness of an absolutely continuous solution to the Cauchy problem
	\begin{equation}\label{cp}
		\left\{
		\begin{aligned}
			(D_m+V) \Psi &= \lambda\Psi\\
			\Psi(0) &= \Phi(0)= \alpha \Xi(0)\,.
		\end{aligned}
		\right.
	\end{equation}
	Therefore, and since $\Phi$ and $\Xi$ are absolutely continuous, we obtain
	\[
		\Phi=\alpha\Xi\,,\quad \text{a.e. on } \R\,.
	\]
	If instead $\Xi(0)=0$, the same uniqueness result implies that $\Xi\equiv0$, which contradicts the fact that $\Xi$ is a generalized eigenfunction.
\end{proof}
\begin{remark}\label{Remark_proof_conditional_simplicity_generalized_eigenvalues}
	The constancy of the Wronskian obtained above is the same mechanism used in the proof of~\cite[Lemma~2.3]{AldRicStoVDB-23}. 
	The observation made here is that this argument only relies on the structural condition on the potential given in Assumption~\ref{Potential_sigma_2_condition}, and therefore applies to a larger class of potentials.
	
	In~\cite{AldRicStoVDB-23}, the vanishing of the Wronskian was deduced from the fact that the functions under consideration were eigenfunctions in~$L^2(\R)$ and hence belonged to $H^1(\R)$, which implies that they vanish at infinity. 
	Here, we replace this property by the assumption that, for a given generalized eigenvalue, all associated generalized eigenfunctions share a common component that vanishes at a same point $x_0\in\R\cup\{\pm\infty\}$.
\end{remark}

\section{Absence of threshold eigenvalues}\label{section_absence_of_threshold_eigenvalues}
We now turn to the proof of Theorem~\ref{main_thm_2_noEV}, namely the absence of threshold eigenvalues for the operator $L_0$. The argument combines two complementary oscillation results.

For $p=1$, the conclusion follows immediately from the existence of threshold resonances together with the simplicity of generalized eigenfunctions at the thresholds (Theorem~\ref{main_thm_3_Resonances} established in the previous section).

First, we show that if $(\varphi_1,\varphi_2)\transp$ is an eigenfunction of~$L_0$ associated with an eigenvalue~$\lambda$ in the essential spectrum of~$L_0$, then both $\varphi_1$ and $\varphi_2$ must vanish infinitely many times on every interval $(R,+\infty)$; see Lemma~\ref{Lemma_L0s_eigenfunction_at_and_above_threshold_vanish_infinitely_many_times}. In particular, the combinations $\varphi_1+\varphi_2$ and $\varphi_1-\varphi_2$ also possess infinitely many zeros on~$(R,+\infty)$.

On the other hand, we prove that $\varphi_1+\varphi_2$ and $\varphi_1-\varphi_2$ satisfy Schr\"odinger equations naturally associated with~$L_0$. Using an oscillation argument inspired by Kneser's approach~\cite{Kneser-1893}, based on the Sturm comparison theorem~\cite[§XII]{Sturm-1836} and a suitable Cauchy--Euler equation, we show that any solution corresponding to the threshold energies $\pm m-\omega$ can have only finitely many zeros on~$\R$; see Lemma~\ref{Lemma_Schr_eigenfunction_at_threshold_vanish_finitely_many_times}. This contradiction excludes the existence of threshold eigenvalues for~$L_0$.

We begin by stating the two lemmas. We then prove Theorem~\ref{main_thm_2_noEV} assuming them, and finally establish each lemma.

\begin{lemma}\label{Lemma_L0s_eigenfunction_at_and_above_threshold_vanish_infinitely_many_times}
	Let $(p,\omega)\in(0,+\infty)\times(0,m)$. Let $(\varphi_1,\varphi_2)\transp$ be an eigenfunction of~$L_0$ associated with an eigenvalue~$\lambda$ satisfying $|\lambda+\omega|\geq m$. Then, the eigenfunction components never vanish simultaneously at a point, i.e. $(\varphi_1,\varphi_2)(x)\neq 0$ for all $x\in \R$. Moreover, 
	\begin{equation}\label{Lemma_eigenfunction_at_and_above_threshold_vanish_infinitely_many_times_eq}
		\forall\, R>0\,, \ \exists\, (x, y)\in (R, +\infty)^2\,, \qquad \varphi_1(x) = 0 = \varphi_2(y)\,.
	\end{equation}
\end{lemma}

Our second ingredient is then the following result.
\begin{lemma}\label{Lemma_Schr_eigenfunction_at_threshold_vanish_finitely_many_times}
	Let $(p,\omega)\in(0,+\infty)\times(0,m)$ and $M$ be defined in~\eqref{Def_M} with $f(s)=s|s|^{p-1}$. If $\lambda\leq m^2$, then any non-trivial solution to $(- \partial_x^2 + M^2 \pm M' -\lambda) y=0$ has finitely many zeros on~$\R$.
\end{lemma}
%Note that, in addition to its use in the proof of Theorem~\ref{main_thm_2_noEV}, \textcolor{blue}{we also make use of it in the proof of Lemma~\ref{Lemma_thresholds_Aq_not_accumulation_points}, which in turn we need in our argument to prove Proposition~\ref{prop-birth} and, consequently, Theorem~\ref{main_thm_1_gap_p_below_1}.}

We now use these two lemmas to prove Theorem~\ref{main_thm_2_noEV}.
\begin{proof}[Proof of Theorem~\ref{main_thm_2_noEV}]
	We first observe that if $\lambda-\omega$ is an eigenvalue of~$L_0$ with eigenfunction $(\varphi_1,\varphi_2)\transp$, then $\lambda$ is an eigenvalue of~$\cA:=L_0+\omega$ with the same eigenfunction. Moreover, $-\lambda$ is an eigenvalue with eigenfunction $\sigma_1(\varphi_1,\varphi_2)\transp=(\varphi_2,\varphi_1)\transp$. Consequently, $\lambda^2$ is a double eigenvalue of~$U \cA^2 U$ with eigenfunctions $(\varphi_1+\varphi_2,0)\transp$ and $(0,\varphi_1-\varphi_2)\transp$, see~\eqref{oldcA}. In particular, $\lambda^2$ is an eigenvalue of both Schrödinger operators $-\partial_x^2+M^2\mp M'$ appearing in the diagonal of $U \cA^2 U$ with respective eigenfunctions $\varphi_1\pm\varphi_2$.
	
	Lemma~\ref{Lemma_L0s_eigenfunction_at_and_above_threshold_vanish_infinitely_many_times} applied to the thresholds $\pm m-\omega$ of~$L_0$ provides increasing sequences $\{x_k\}_{k\in\N}$ and $\{y_k\}_{k\in\N}$ diverging to $+\infty$ such that
	\[
		\forall\, k\in\N\,, \qquad \varphi_2(x_k)\neq\varphi_1(x_k)=0=\varphi_2(y_k)\neq\varphi_1(y_k)\,.
	\]
	For the function $h:=\varphi_1^2-\varphi_2^2$ this yields
	\[
		\forall\, k\in\N\,,\quad
		\left\{
		\begin{aligned}
			h(x_k)&=\varphi_1^2(x_k)-\varphi_2^2(x_k)=-\varphi_2^2(x_k)<0\,,\\
			h(y_k)&=\varphi_1^2(y_k)-\varphi_2^2(y_k)=\varphi_1^2(y_k)>0\,.
		\end{aligned}
		\right.
	\]
	Since $h$ is continuous (Proposition~\ref{bas1}), it therefore admits infinitely many zeros on any interval $(R,+\infty)$.
	
	We now claim that both $\varphi_1+\varphi_2$ and $\varphi_1-\varphi_2$ consequently have infinitely many zeros on any interval $(R,+\infty)$. Since the argument for~$\varphi_1+\varphi_2$ is identical (after replacing $\varphi_2$ by $-\varphi_2$), we only present the proof for~$\varphi_1-\varphi_2$. Assume by contradiction that $\varphi_1-\varphi_2$ has only finitely many zeros on~$(R,+\infty)$ for some $R>0$. Up to replacing $\varphi_i$ by $-\varphi_i$ and increasing $R$, we may assume without loss of generality that $\varphi_1>\varphi_2$ on~$(R,+\infty)$.
	
	This implies $\varphi_1>0>\varphi_2$ on~$(R,+\infty)$, contradicting~\eqref{Lemma_eigenfunction_at_and_above_threshold_vanish_infinitely_many_times_eq}. Indeed, if $\varphi_1\le0$ somewhere on~$(R,+\infty)$, then $\varphi_2<0$, hence $\varphi_1+\varphi_2<0$ and $h=\varphi_1^2-\varphi_2^2<0$, contradicting that $h$ has infinitely many zeros on~$(R,+\infty)$. Similarly, if $\varphi_2\ge0$ somewhere on~$(R,+\infty)$, then $\varphi_1>0$, which yields the same contradiction.
	
	To summarize, we have shown that $\varphi_1\pm\varphi_2$ have infinitely many zeros while being eigenfunctions of the Schrödinger operators $-\partial_x^2+M^2\mp M'$ associated with the eigenvalue $m^2$. Lemma~\ref{Lemma_Schr_eigenfunction_at_threshold_vanish_finitely_many_times} yields the desired contradiction, proving that the thresholds $\pm m-\omega$ cannot be eigenvalues of~$L_0$.
\end{proof}

We now turn to the proofs of the two lemmas. The proof of Lemma~\ref{Lemma_L0s_eigenfunction_at_and_above_threshold_vanish_infinitely_many_times} relies on the continuity of~$\Psi$ ensured by Proposition~\ref{bas1}, which holds for any generalized eigenvalue $\lambda\in\R$ and any potential $V\in L^1_{\rm loc}(\R)$.

Moreover, the argument applies to the Soler model with a much broader class of nonlinearities than the power nonlinearity $f(s)=s|s|^{p-1}$. In fact, it works for any nonlinearity $f$ satisfying the generic assumptions introduced in~\cite[Assumption~1.1]{AldRicStoVDB-23}. These assumptions ensure the existence of the solitary wave $\phi_0$ and, through the properties of~$\phi_0$ (see, e.g.,~\cite[Proposition~2.1]{AldRicStoVDB-23}), imply that
\[
	V=-f\!\left(\pscal{\phi_0}{\sigma_3\phi_0}_{\Cbb^2}\right)\sigma_3 \in L^\infty(\R,\Cbb^{2\times2})\,.
\]
For convenience we recall the relevant part of these assumptions.

\begin{assumption}[First part of~{\cite[Assumption~1.1]{AldRicStoVDB-23}}]\label{Assumption_existence_solitary_waves}
	$f\in \mathcal{C}^1(\R\setminus\{0\},\R)\cap\mathcal{C}^0(\R,\R)$ with $f(0)=0$, $\lim_{+\infty}f\geq m$, and $f'>0$ on~$(0,+\infty)$.
\end{assumption}

We therefore prove Lemma~\ref{Lemma_L0s_eigenfunction_at_and_above_threshold_vanish_infinitely_many_times} under the more general Assumption ~\ref{Assumption_existence_solitary_waves}, and for any eigenvalue $\lambda$ of~$L_0$ such that $|\lambda+\omega|\geq m$.

For notational convenience, the proof is written in terms of the operator $\cA=L_0+\omega$. Thus, $\lambda$ is considered as an eigenvalue of~$\cA$, and the assumption becomes $\lambda\equiv\lambda_{\cA}\geq m$.

\begin{proof}[Proof of Lemma~\ref{Lemma_L0s_eigenfunction_at_and_above_threshold_vanish_infinitely_many_times}]
	Let $\lambda\geq m$ be an eigenvalue of~$\cA=L_0+\omega$ with associated eigenfunction $(\varphi_1,\varphi_2)\transp$. Since the eigenvalues of~$\cA$ are simple~\cite[Lemma~2.3]{AldRicStoVDB-23} and $\cA$ preserves $L^2_\pm(\R,\R^2)$, separating real and imaginary parts shows that both are eigenfunctions as well. Hence the eigenfunction is real-valued up to a complex constant, and we may assume without loss of generality that $(\varphi_1,\varphi_2)$ is real-valued.

	The eigenvalue equation for~$\cA$ reads
	\begin{equation}\label{EV_equation_lambda}
		\left\{
		\begin{aligned}
			\varphi_2' &= (\lambda-M)\varphi_1\,,\\
			\varphi_1' &= -(\lambda+M)\varphi_2\,,
		\end{aligned}
		\right.
	\end{equation}
	where $M$ is defined in~\eqref{Def_M}. As mentioned before, by~\cite[Proposition~2.1]{AldRicStoVDB-23}, Assumption~\ref{Assumption_existence_solitary_waves} implies that $V=-f\!\left(\pscal{\phi_0}{\sigma_3\phi_0}_{\Cbb^2}\right)\in L^\infty(\R) \subset L^1_{\rm loc}(\R)$ and Proposition~\ref{bas1} therefore yields $(\varphi_1,\varphi_2)\transp\in \ac(\R)$ and the eigenfunction is in particular continuous.

	We also note that $(\varphi_1,\varphi_2)(x)\neq 0$ for all $x\in\R$, as claimed. Indeed, if there existed $x_0\in\R$ such that $(\varphi_1,\varphi_2)(x_0)=0$, then~\eqref{EV_equation_lambda} would also give $(\varphi_1',\varphi_2')(x_0)=0$. Since $(\varphi_1,\varphi_2)\transp$ solves a first-order ODE, vanishing together with its derivative at $x_0$ would imply that it is identically zero, contradicting that it is an eigenfunction, see~Proposition~\ref{bas1}.
	
	We conclude with the proof of~\eqref{Lemma_eigenfunction_at_and_above_threshold_vanish_infinitely_many_times_eq}. That is,
	\[
		\forall\, R>0\,, \ \exists\, (x, y)\in (R, +\infty)^2\,, \qquad \varphi_1(x) = 0 = \varphi_2(y)\,.
	\]
	Since the argument is the same for~$\varphi_1$ and $\varphi_2$, we present it only for~$\varphi_1$. Assume by contradiction that the property~\eqref{Lemma_eigenfunction_at_and_above_threshold_vanish_infinitely_many_times_eq} does not hold for~$\varphi_1$. Since $(\varphi_1,\varphi_2)\transp$ is continuous, there exists $R>0$ such that $\varphi_1>0$ on~$(R,+\infty)$ or $-\varphi_1>0$ on~$(R,+\infty)$. Without loss of generality (replacing $(\varphi_1,\varphi_2)\transp$ by $(-\varphi_1,-\varphi_2)\transp$ if necessary) we assume $\varphi_1>0$ on~$(R,+\infty)$.
	
	Noticing first that Assumption~\ref{Assumption_existence_solitary_waves} actually implies that $M<m$ on~$\R$ with $M(x)\to m$ as~$|x|\to+\infty$, we deduce for~$\lambda\geq m$ that the quantities $\lambda-M$ and $\lambda+M$ appearing in~\eqref{EV_equation_lambda} are both positive.

	Therefore, for~$R$ sufficiently large, $\varphi_1>0$ on~$(R,+\infty)$ together with equation~\eqref{EV_equation_lambda} and the positivity of~$\lambda-M$ imply that $\varphi_2'>0$ on~$(R,+\infty)$. This in turn yields $\varphi_2<0$ on~$(R,+\infty)$, since otherwise there would exist $r\geq R$ such that $\varphi_2(r)\ge0$ while $\varphi_2'>0$ on~$(r,+\infty)$, contradicting the fact that $\varphi_2\in L^2(\R)$.
	
	Taking $R$ larger if necessary, $\varphi_2<0$ on~$(R,+\infty)$ together with equation~\eqref{EV_equation_lambda} and the positivity of~$\lambda+M$ implies $\varphi_1'>0$ on~$(R,+\infty)$. This concludes the proof as it contradicts $\varphi_1\in L^2(\R)$, since we would have $\varphi_1>0$ and $\varphi_1'>0$ on~$(R,+\infty)$.
\end{proof}

We now prove Lemma~\ref{Lemma_Schr_eigenfunction_at_threshold_vanish_finitely_many_times}. Our argument relies on the fast decay at $\pm\infty$ of the potentials $M^2-m^2\pm M'$ appearing in the Schrödinger operators derived above. The relevant case is $\lambda=m^2$; however, for completeness we state and prove the result for all $\lambda\leq m^2$.

\begin{proof}[Proof of Lemma~\ref{Lemma_Schr_eigenfunction_at_threshold_vanish_finitely_many_times}]
	As recalled in the proof of Lemma~\ref{Lemma_L0s_eigenfunction_at_and_above_threshold_vanish_infinitely_many_times}, under Assumption~\ref{Assumption_existence_solitary_waves}—and in particular for the power nonlinearities $f(s)=s|s|^{p-1}$—we have $M^2-m^2<0$ and $M^2(x)-m^2\to0$ as $|x|\to\infty$. 
	
	Moreover, for power nonlinearities one has (see, e.g.,~\cite[(27)]{AldRicStoVDB-23})
	\[
		M(x)=m-2m(p+1)\frac{\nu}{1+\nu} \frac{1-\tanh^2(p\kappa x)}{1-\nu\tanh^2(p\kappa x)}\,,
	\]
	where $\kappa=\sqrt{m^2-\omega^2}$ and $\nu=(m-\omega)/(m+\omega)$. A direct computation gives
	\[
		M'(x)=4m(p+1)\nu\frac{1-\nu}{1+\nu} \frac{1-\tanh^2(p\kappa x)}{1-\nu\tanh^2(p\kappa x)} \frac{\tanh(p\kappa x)}{1-\nu\tanh^2(p\kappa x)}\,.
	\]
	Hence $M(x)\nearrow m$ and $M'(x)\to0$ exponentially fast as $x\to\pm\infty$.
	
	Suppose first that $\lambda<m^2$. Applying the Sturm comparison theorem to
	\[
		q_1:=-(M^2-\lambda\pm M') \quad \text{ and } \quad q_2:=0\,,
	\] we observe that $q_1<q_2$ on~$(-\infty,-R)\cup(R,+\infty)$ for some $R>0$. Since any non-trivial solution to $y'' + q_2y = y'' =0$ has at most one zero, it follows that any non-trivial solution to $y'' + q_1 y=0$ has at most two zeros on~$(-\infty,-R)\cup(R,+\infty)$.

	The function $q_1$ being continuous, there exists $C>0$ such that $q_1\leq C$ on the compact interval $[-R,R]$. Since solutions of~$y''+Cy=0$ have finitely many zeros on~$[-R,R]$, another application of the Sturm comparison theorem shows that the same holds for any non-trivial solution to $y'' + q_1 y=0$. This proves the lemma when $\lambda<m^2$.
	
	We now consider the case $\lambda=m^2$. In this case $q_1$ is no longer necessarily negative on both $(-\infty,-R)$ and $(R,+\infty)$, so the previous argument does not apply directly. However, the exponential decay established above implies $x^2q_1(x)\to0$ as $x\to\pm\infty$. Hence, Kneser's criterion~\cite{Kneser-1893} is satisfied. Namely, $x^2q_1(x)<1/4$ for large $|x|$. It follows that there exists $R>0$ such that any non-trivial solution to $y'' + q_1 y=0$ with $\lambda=m^2$ has finitely many zeros on~$(-\infty,-R)\cup(R,+\infty)$. 
	
	The argument on the compact interval $[-R,R]$ is the same as above, and the proof is complete.
\end{proof}

%\appendix
%    \phantomsection %anchor: for the link in TOC created by the next by next instruction to send to the right page.
%    \addcontentsline{toc}{section}{Appendices}
%    \addtocontents{toc}{\protect\setcounter{tocdepth}{0}} % To avoid automatic additions of the section "Appendix" and its subsections in the TOC.

\bibliographystyle{siam}
%\bibliography{/Users/j/Dropbox/LaTeX/biblio.bib}
\newcommand{\noop}[1]{}

\end{document}